\documentclass[11pt]{article}

\usepackage[T1]{fontenc}
\usepackage{lmodern}
\usepackage[margin=1in]{geometry}
\usepackage{graphicx}
\usepackage{amsmath,amssymb,amsfonts}
\usepackage{bm}
\usepackage{subcaption}
\usepackage{float}
\usepackage{adjustbox}
\usepackage{placeins}
\usepackage{gensymb}
\usepackage{booktabs}
\usepackage{xurl}
\usepackage{hyperref}
\hypersetup{colorlinks=true,allcolors=blue}
\usepackage[
    backend=biber,
    style=authoryear,
    maxcitenames=2,
    maxbibnames=99,
    uniquelist=false
]{biblatex}
\newcommand{\papertitle}{%
Mapping urban heat from citizen-collected data:\\[0.12em]
a scalable Gaussian process approach to\\[0.12em]
correcting for preferential sampling}

\begin{document}

\begin{center}
\vspace*{-0.35in}

{\fontsize{18.8}{22.5}\selectfont\bfseries
\papertitle\par}

\vspace{1.25em}

{\large
Ellie K. Kim$^{1}$,
Zachary D. Calhoun$^{2}$,
Michael H. Bergin$^{2}$,\\[0.15em]
David E. Carlson$^{1,2,3}$\par}

\vspace{0.9em}

{\small
$^{1}$Department of Computer Science, Duke University, USA\\
$^{2}$Department of Civil and Environmental Engineering, Duke University, USA\\
$^{3}$Department of Biostatistics and Bioinformatics, Duke University, USA\par}

\vspace{0.75em}

{\small
Corresponding author: Ellie Kim;
\href{mailto:ellie.kim@duke.edu}{ellie.kim@duke.edu}\par}

\end{center}

\vspace{1.1em}

\begin{abstract}
The urban heat island (UHI) effect amplifies heat stress in cities, increasing heat-related illness, mortality, sleep disruption, and energy demand. Because UHI intensity varies at neighborhood scales, intra-city temperature mapping is essential for assessing public health impacts, yet existing monitoring networks are sparse. Crowdsourced weather stations provide dense, hourly observations but suffer from noise and systematic biases, including preferential sampling driven by where residents install sensors. We introduce two scalable Gaussian process frameworks which adjust for preferential sampling and enable statewide, high-resolution temperature prediction. The first integrates sampling correction directly within a sparse variational Gaussian process (SVGPR+PS). The second uses a two-stage estimator involving a log-Gaussian Cox process (LGCP+GP) that estimates the sampling intensity field and incorporates it into temperature prediction. Both methods rely on inducing-point approximations and variational inference to scale beyond traditional Gaussian processes. Applied to approximately 10,000 citizen stations across North Carolina, the models generate hourly temperature fields within minutes. Validation against independent urban heat campaign data shows that the preferential sampling correction improves predictive performance, increasing average afternoon correlations across four cities from 0.26 to 0.43. This scalable framework enables neighborhood-level UHI assessment and supports studies of heat exposure disparities statewide.
\end{abstract}


\section*{Impact Statement}
While crowdsourced weather data is essential for neighborhood-level mapping, sensor placement is often biased. Personal weather stations tend to be located in wealthier, cooler neighborhoods, causing standard models to underestimate heat in vulnerable communities. We present a scalable statistical method that corrects for this bias, generating hourly block-group temperature maps for the entire state of North Carolina. Our corrected estimates significantly improve agreement with independent ground-truth observations compared to both uncorrected models and existing reanalysis products. These bias-corrected temperature fields enable more detailed heat exposure assessment. Specifically, they allow for public health interventions during extreme heat events as well as epidemiological studies pairing neighborhood-level heat exposure with hospitalization records, mortality data, and sleep quality metrics.
 
\begin{figure}[H]
    \centering
    \includegraphics[width=0.75\linewidth]{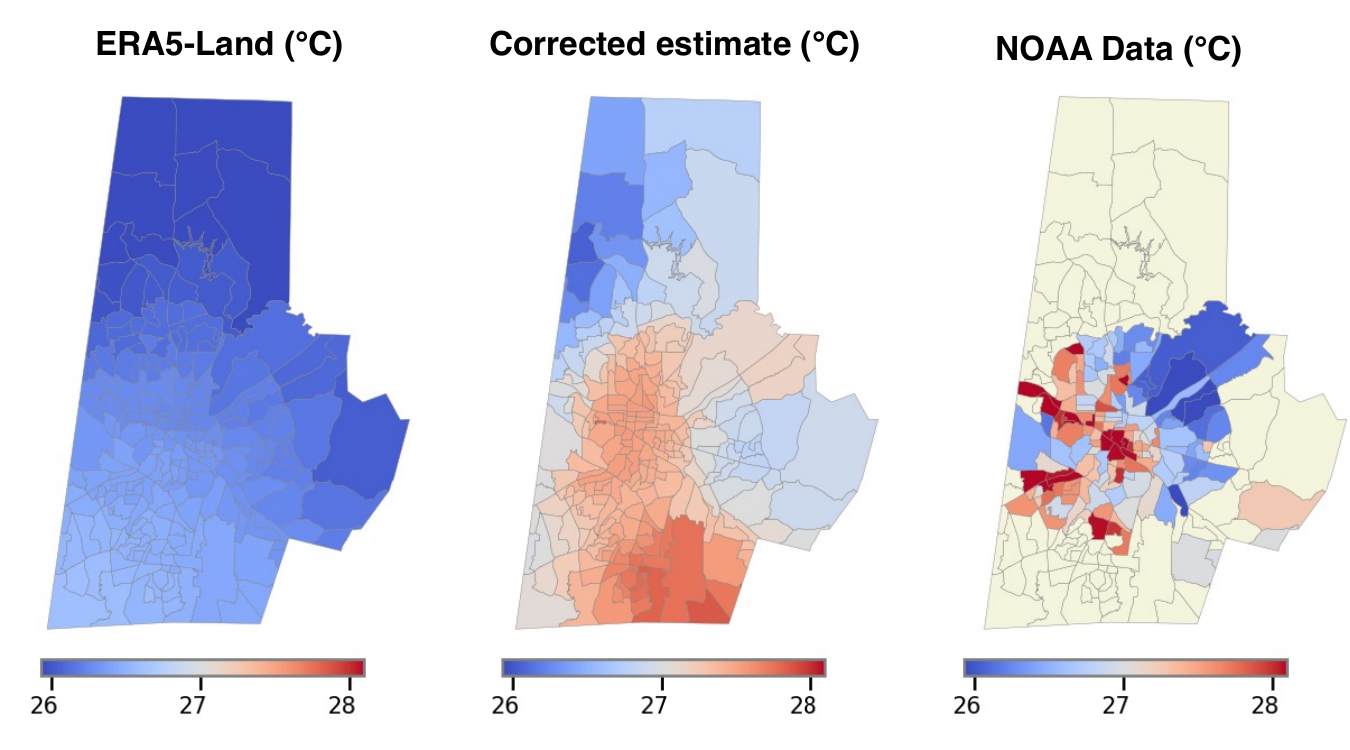}
    \caption{Comparison of ERA5-Land, our LGCP+GP corrected estimate, and NOAA urban heat island campaign data for July 23, 2021 (7--8 P.M.) block groups in Durham, North Carolina. ERA5-Land was upsampled via bilinear interpolation then aggregated to units. The corrected estimate resolves neighborhood-scale variability that is less apparent in ERA5-Land and more consistent with NOAA observations.}
    \label{fig:overview_comparison}
\end{figure}
 
\section{Introduction}
 
The urban heat island (UHI) effect is a phenomenon in which urban areas experience higher temperatures than surrounding rural regions due to built infrastructure and human activity \parencite{Oke}. For example, vegetation may reflect solar radiation while materials like concrete and asphalt absorb and trap heat. Elevated urban temperatures worsen heatwaves and disproportionately affect vulnerable populations \parencite{Khatana}. They also elevate demand for air conditioning, which increases energy consumption and greenhouse gas emissions \parencite{santamouris_2020_HeatMitigationTechnologies}. Therefore, accurate fine-scale temperature modeling is essential for heat mitigation planning and public health intervention. For example, neighborhood-level temperature estimates can inform cooling center placement, environmental justice screening, and epidemiological studies linking heat exposure to health outcomes.
 
A major barrier to predicting air temperature at the neighborhood level is the lack of high-resolution urban temperature measurements. For example, traditional meteorological stations tend to be located outside of urban centers and there may be only a few of them to represent an entire city. Mobile transects, such as NOAA's urban heat mapping efforts \parencite{NOAA}, are a high-resolution alternative. In this form of data collection, volunteers equip cars, bikes, or handheld sensors to measure temperature along predefined routes. These traversal data naturally have high spatial coverage; however, they only provide data for the short time intervals during which the traversals occurred. On the other hand, weather station networks provide frequent measurements, but tend to be expensive, causing them to be sparse. For example, North Carolina State University's ECONet consists of 44 sensors across all of North Carolina \parencite{ECONet}. Furthermore, reanalysis products such as ERA5-Land provide hourly measurements on a fairly dense 9-km grid \parencite{essd-13-4349-2021}, but their dependence on sparse observations limits their ability to capture UHI effects at the neighborhood level.
 
Thus, citizen-collected data from sources like Weather Underground (WU) are a promising solution, as they provide high spatial and temporal resolution through low-cost personal weather stations (PWS). WU data are collected via stations installed by citizen volunteers. Due to ease of setup, the WU network is much denser than traditional weather station networks. Additionally, these stations are able to record near-surface meteorological variables such as air temperature, humidity, pressure, and precipitation at intervals ranging from 1 to 5 minutes.
 
Despite their advantages, citizen-collected data are not without limitations; they are prone to noise and systematic bias. For example, citizen observations are often affected by \textit{preferential sampling}, where the likelihood of collecting a sample at a specific location is influenced by the value of the variable being studied \parencite{Diggle}. This occurs because residents are more likely to install weather stations in their own neighborhoods, where urban heat islands tend to be less prevalent. Consequently, studies utilizing citizen data may underestimate neighborhood heat exposure, masking the very disparities UHI studies aim to highlight \parencite{calhoun2024refining}.
 
Preferential sampling can be addressed by adjusting temperature estimates based on observation density. Previous studies have accounted for preferential sampling by defining the estimated temperature in relation to the underlying spatial process driving variation \parencite{Diggle}. This process is commonly modeled as a Gaussian process (GP), a precedent we follow here. However, given that GPs tend not to scale well beyond a few thousand data points, it has become common in recent years to use an approximation. Prior work by \textcite{calhoun2024refining} used an intrinsic conditional autoregressive (ICAR) model and achieved considerable success when their model was applied to Durham County in North Carolina. However, this study relied on monthly averaged data in order to avoid short-term meteorological variability obscuring the spatial structure of temperature patterns. As a result, predictions could only be made on a monthly basis \parencite{calhoun2024refining}.
 
In this article, we present two approaches to combat issues of preferential sampling and GP scalability. The first incorporates the preferential sampling correction into a sparse variational GP (SVGPR+PS). The second incorporates inferred sampling intensity into a second GP for temperature prediction (LGCP+GP). The LGCP+GP approach is somewhat related to the method proposed by \textcite{hsiao2025inverse}, in that we separate the preferential sampling adjustment into a two-stage estimator. However, rather than substituting the sampling intensity estimator for a kernel density estimator, we approximate a log-Gaussian Cox process with variational inference. Both methods leverage inducing-point approximations and stochastic variational inference to scale to statewide datasets comprising tens of thousands of locations.
 
We validate our models against independent NOAA UHI campaign data \parencite{NOAA}, comparing predictions to machine-learned raster products derived from mobile transects. Additionally, we compare our approach with ERA5-Land, a widely used reanalysis product that performs well at capturing statewide temperature patterns but operates at a relatively coarse spatial resolution (9 km) \parencite{essd-13-4349-2021}. This comparison demonstrates that our methods are better able to capture localized temperature differences, thereby improving heat exposure assessment, which has important implications for epidemiological research and public health policy.
 
\section{Related work}
 
Citizen science data provide high spatial and temporal resolution for environmental monitoring but raise concerns about sampling bias. In spatial statistics, this issue is formalized as preferential sampling, which occurs when observation locations depend on the underlying process of interest. \textcite{Diggle} provide the foundational statistical framework for preferential sampling. Specifically, they introduce a shared latent process (SLP) model in which sampled locations and observed values depend on the same underlying spatial field. They also demonstrate that ignoring this dependence can lead to biased parameter estimates and misleading spatial predictions.
 
Prior work has demonstrated that the WU temperature data are preferentially sampled. For example, \textcite{calhoun2024refining} demonstrated that the WU air temperature data in Durham County, North Carolina, are preferentially sampled with respect to neighborhood income. The study found that WU stations are disproportionately concentrated in higher-income areas, while lower-income neighborhoods (where urban heat risks are often greatest) are under-observed. We extend these findings by demonstrating statistically significant preferential sampling across all of North Carolina.
 
Adjusting for preferential sampling in citizen weather data has been accomplished in prior work. In particular, \textcite{calhoun2024refining} proposed a Bayesian framework to correct for preferential sampling in WU air temperature data, which follows the SLP approach \parencite{Diggle}. Their analysis provided important evidence that citizen weather stations are disproportionately concentrated in higher-income neighborhoods and that accounting for this bias substantially alters model inference. However, the study was limited in several key respects. First, inference was only conducted across a single county. Second, model fitting and validation relied on temporally averaged monthly temperature data, in order to reduce noise from synoptic variability and to make inference tractable under Markov chain Monte Carlo (MCMC). Thus, their approach does not readily scale to hourly data. Our work addresses these limitations by separately learning the sampling intensity field, enabling inference at finer temporal resolution without requiring temporal averaging. We model the spatial process of interest with a sparse variational Gaussian process rather than a conditional autoregressive model, allowing the method to scale to statewide hourly data.
 
Finally, we note that Bayesian inference for log-Gaussian Cox processes and preferential sampling is often performed using Integrated Nested Laplace Approximation (INLA). In this work, we instead adopt variational inference, which has native support in Python (i.e.\ Pyro), in order to enable seamless integration into future methods.
 
\section{Methods and materials}
 
\subsection{Data collection}
 
Our analysis integrates multiple data sources to model air temperature across North Carolina. We combine observations from WU with environmental covariates such as elevation and land surface temperature (LST), as well as demographic information from the U.S.\ Census.
 
\subsubsection{Temperature datasets}
 
We used temperature observations from Weather Underground, a citizen-science platform that aggregates data from personal weather stations. The WU network provides high spatiotemporal resolution through thousands of low-cost personal weather stations installed by citizen volunteers. For example, as of 2025, Durham County hosts over 300 stations. These stations are distributed across both urban and suburban areas. We retrieved air temperature measurements for all North Carolina stations via the WU API, retaining only values passing internal quality control standards. These measurements were then filtered to three time slots corresponding to NOAA's UHI campaign data-collection windows: morning (6--7 a.m.), afternoon (3--4 p.m.), and evening (7--8 p.m.). All times are given in Eastern Daylight Time (EDT). Measurements were additionally filtered to the last ten minutes of every hour to ensure comparisons with hourly data (i.e., ERA5-Land) would be possible.
 
For each station, we retained a single representative temperature per slot: the median of all readings from that station. The median was used rather than an alternative metric such as the mean, in order to reduce the influence of station-specific noise and outliers.
 
\subsubsection{Covariates}
 
Elevation is a widely-used covariate in atmospheric and climate modeling due to the adiabatic lapse rate. Our elevation data are derived from the U.S.\ Geological Survey (USGS) digital elevation model (DEM). These data were aggregated to the areal units used in our analysis (census tracts and block groups). Specifically, we computed the mean DEM value within each areal unit.
 
Additionally, land surface temperature (LST) was used as a covariate due to its ability to reflect how different surfaces absorb and retain heat. Our LST data are sourced from NASA's Moderate Resolution Imaging Spectroradiometer (MODIS) aboard the Terra and Aqua satellites. MODIS observes the entire Earth's surface every 1 to 2 days, collecting data in 36 spectral bands at spatial resolutions of 250, 500, and 1,000 meters. We used daytime and nighttime surface temperature observations derived from these bands, which are publicly distributed through the Land Processes DAAC by the U.S.\ Geological Survey and were accessed via the Google Earth Engine platform \parencite{GEE}. To construct a robust estimate of average LST for each month, we downloaded daily satellite images and computed the pixel-wise mean over the course of a month. This yielded a composite raster representing the average daytime or nighttime surface temperature over North Carolina for that month. To ensure accurate aggregation for extremely small spatial units, we resampled the MODIS raster to a resolution ten times finer than the original using bilinear interpolation before taking the average.
 
Prior to fitting the models, missing elevation and LST values were imputed using a k-nearest neighbors approach with distance weighting, ensuring that missing covariates were filled based on nearby regions. Additionally, we excluded water regions from our analysis. All covariates were standardized to a mean of 0 and standard deviation of 1 following imputation.
 
Lastly, we incorporated demographic covariates from the American Community Survey, an annual survey that is part of the U.S.\ Census Bureau's Decennial Census Program \parencite{ACS}. Specifically, population and median household income data were obtained on a census tract and block group basis.
 
\section{Statistical Models}
 
We propose two complementary models for estimating air temperature while accounting for preferential sampling. The first model, SVGPR+PS, incorporates the preferential sampling adjustment directly into a sparse variational Gaussian process (SVGP). A shared latent GP drives sampling intensity as well as variations in temperature. The second model, LGCP+GP, separates these two processes. It first models the spatial locations of weather sensors as a Poisson process driven by a latent GP. The inferred intensity field is then incorporated into a separate GP used to predict temperature. We first introduce the SVGPR+PS model, then present LGCP+GP as an alternative framework providing additional flexibility. Note that more information on SVGPs can be found in the appendix.
 
Figure~\ref{fig:svgpr_ps_vs_lgcp_gp} illustrates the structure of both models. Throughout the remainder of this section, scalars are denoted by lowercase non-bold symbols, vectors by lowercase bold symbols, matrices by uppercase bold symbols, and functions by lowercase non-bold symbols.
 
\begin{figure}[t]
    \centering
    \includegraphics[width=0.96\textwidth]{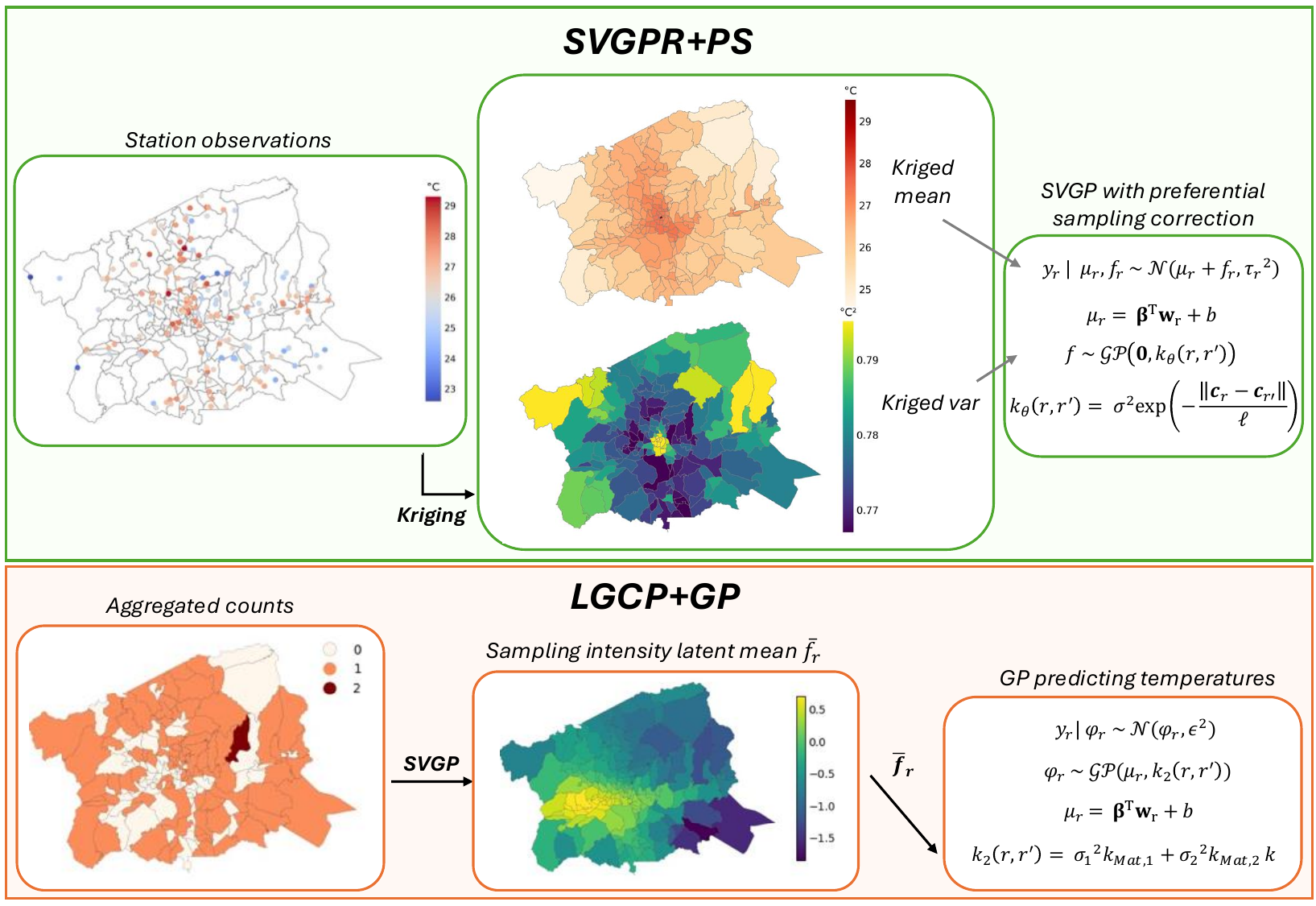}
    \caption{Model summaries for SVGPR+PS and LGCP+GP.}
    \label{fig:svgpr_ps_vs_lgcp_gp}
\end{figure}
 
\subsection{Shared latent process model}
 
In order to adjust for preferential sampling, we borrow heavily from \textcite{Diggle}'s shared latent process (SLP) model (background information on this original model is found in the appendix). Note that the original SLP model \parencite{Diggle} assumed constant mean, a single shared latent process, and a Gaussian likelihood. We deviate from this by adding covariates into the mean and using an SVGP rather than an exact GP, as inspired by \textcite{pati2011bayesian} and \textcite{titsias_2009} correspondingly. We also transition this model from the point-wise setting into the purely areal setting, which is more relevant in the context of spatial UHI modeling.
 
\subsection{Model 1: SVGPR+PS}
 
In the SVGPR+PS model, we formulate the preferential sampling mechanism at the areal level. Specifically, we approximate the LGCP by modeling station counts aggregated within each census unit rather than at exact point locations. For an inhomogeneous Poisson process, the expected number of events in region $A$ can be expressed as the integral of the intensity function $\lambda(\cdot)$ over that region: $\int_A \lambda(\mathbf{x}) d\mathbf{x}$ \parencite{kingman1993poisson}. When the intensity surface is smooth, this integral can be reasonably approximated by averaging the value of the intensity function at a set of representative points called quadrature points \parencite{berman1992approximating}. In our setting, we use the set of all centroids $\mathcal{C}$ as our representative points and scale intensity $\lambda$ of each areal unit by its corresponding population. This creates a Poisson process across areal units rather than individual data points.
 
We begin with point-wise temperature observations from WU stations. To translate this data to the areal setting, we use a GP regression model to interpolate the temperature field and obtain a single value for each areal unit. We use the posterior mean and variance to fit the subsequent model.
 
\subsubsection{Preferential sampling adjustment}
 
In the following, we index areal units with $r = 1, ..., R$, where the $R$ units partition the spatial domain $\mathcal{D} \subset \mathbb{R}^2$ corresponding to the entire state of North Carolina. Note that these areal units do not overlap. Let $\mathbf{c}_r \in \mathcal{C}$ denote the centroid of areal unit $r$, where $\mathcal{C}$ is the set of all centroids in $\mathcal{D}$. Lastly, let $n$ be the total number of observed stations from WU.
 
Borrowing from \textcite{Diggle}, it is customary to model the shared latent process $f$ as a GP; in our case, we use a sparse variational Gaussian process (SVGP) for improved scalability. In the following, we abbreviate $f(\mathbf{c}_r)$ as $f_r$. In the mean module, $\mathbf{w}_r$ is the vector of standardized covariates corresponding to unit $r$, and $\boldsymbol{\beta}$ and $b$ are the associated constants for these covariates. $y_r$ is the value of the temperature field at centroid $\mathbf{c}_r$, while $\tau_r^2$ represents the uncertainty associated with this value.
\begin{align}
y_r \mid \mu_r, f_r &\sim \mathcal{N}(\mu_r + f_r, \tau_r^2), \\
\mu_r &= \boldsymbol{\beta}^\top \mathbf{w}_r + b, \\
f &\sim \mathcal{GP}(\mathbf{0}, k(\cdot, \cdot)), \\
k(r, r') &= \sigma^2 \text{exp} \Big( -\frac{\|\mathbf{c}_r - \mathbf{c}_{r'}\|}{\ell} \Big).
\end{align}
 
The SVGP kernel $k(\cdot, \cdot)$ is an exponential kernel where $\sigma^2$ is the marginal variance and $\ell$ is the length-scale. Note also that we define the covariance over areal units via their centroid coordinates, so $k(\mathbf{c}_r, \mathbf{c}_r')$ is abbreviated as $k(r, r')$.
 
The SVGP approximates the full GP posterior using a set of $M$ inducing points. Inducing locations are initialized via k-means clustering over the latitude--longitude coordinates of centroids in $\mathcal{C}$. Because census units contain approximately equal population, areal units tend to be smaller in dense urban regions. This results in more centroids and thus more inducing points in urban areas.
 
We use an areal approximation to a Poisson process to model sampling intensity. Let $\lambda(\mathbf{x})$ denote the sampling intensity at point location $\mathbf{x}$. Under an areal approximation, we assume the intensity is approximately constant within region $r$, so that:
\begin{align}
    \lambda_r = \text{exp}(\alpha_0 + \alpha_1 f_r) \,\,\frac{p_r}{A_r}, \\
    \int_{A_r} \lambda(\mathbf{x}) \, d\mathbf{x} \approx \lambda_r \, A_r, \\ \label{eq:integral_approx}
    \alpha_0 \sim \mathcal{N}(-1.0,\,1.0), \\
    \alpha_1 \sim \mathcal{N}(-0.5,\,0.5).
\end{align}
 
$f_r$ is the same SLP modeled by the SVGP evaluated at $\mathbf{c}_r$, and $\lambda_r$ is the sampling intensity per unit area of region $r$. $\alpha_0$ and $\alpha_1$ are coefficients controlling baseline intensity and preferential sampling strength correspondingly. The informative priors on $\alpha_0$ and $\alpha_1$ were chosen to reflect plausible ranges for baseline sampling intensity and preferential sampling strength while stabilizing optimization. Specifically, the $\alpha_0$ prior centers the baseline intensity at a moderate level, while the $\alpha_1$ reflects a prior belief that preferential sampling effects are not extreme.
 
Note that population density $p_r / A_r$ is incorporated into the intensity function. As we demonstrate in Section 5.1, this inclusion is appropriate for preferentially-sampled citizen science data, where sampling intensity is expected to depend on population $p_r$ \parencite{calhoun2024refining}. Population density is formally defined as the population of region $r$ divided by its area $A_r$. Note that this is the same as simply scaling by population alone; we write this in terms of population density to make the integral approximation in Equation~\ref{eq:integral_approx} more apparent.
 
Let $s_r$ be the number of sampled locations in unit $r$. In the case of the WU data, this is the number of active stations in $r$ during the specified time window. We can now define the inhomogeneous Poisson process in terms of $\lambda_r$:
\begin{align}
s_r \mid f &\sim \text{Poisson} (\lambda_r \, A_r ).
\end{align}
 
A full derivation of the resulting log likelihood is found in the appendix.
 
\subsection{Model 2: LGCP+GP}
 
While SVGPR+PS joins the temperature field and sampling intensity through an SLP, this joint formulation can make it difficult to separate the two processes and to diagnose how much each contributes to prediction accuracy. To fix this issue we present an alternative model: the LGCP+GP model, which is based on ideas initially proposed by \textcite{reich_2012_AccountingDesignAnalysis}. (More information can be found in Appendix~\ref{sec:appendix_lgcp_gp_equivalence}.) The LGCP+GP first learns a spatial intensity field that explains where stations tend to be located, then uses this field to inform the covariance structure of a separate GP modeling temperature. LGCP+GP also offers additional flexibility by allowing the second--stage temperature GP to have its own kernel. Note that this is not a strict log-Gaussian Cox point process. Instead, we are using aggregated counts over a defined area.
 
\subsubsection{Log-Gaussian Cox process}
 
For areal unit $r$ with population $p_r$ and observed station count $s_r$:
\begin{align}
s_r \mid f &\sim \mathrm{Poisson}\,\big(p_r\,\lambda_r\big), \\[4pt]
\log \lambda_r &= \alpha_0 + \alpha_1\,f_r, \\[4pt]
f &\sim \mathcal{GP}\!\big(0,\,k(r,r')\big).
\end{align}
 
This has the same areal Poisson form as SVGPR+PS, but here $\lambda_r$ is interpreted as per-capita station intensity. We take $k(r, r')$ to be the Mat\'ern kernel over spatial coordinates:
\begin{align}
k(r,r')
&= \sigma^2
\left( 1 + \frac{\sqrt{3}\, \|\mathbf{c}_r - \mathbf{c}_{r'}\|}{\ell} \right)
\exp\!\left( -\frac{\sqrt{3}\,\|\mathbf{c}_r - \mathbf{c}_{r'}\|}{\ell} \right).
\end{align}
 
We use a Mat\'ern kernel in the LGCP to impose smoother latent spatial intensity surfaces, whereas SVGPR+PS utilizes an exponential kernel to allow sharper localized spatial variation.
 
We fit this with an SVGP with $M$ inducing points. This yields the same Poisson log-likelihood as SVGPR+PS, except with expected count $p_r \lambda_r$. For numerical stability we scale $p_r$ by $10^{-3}$ and add a small offset of $10^{-2}$.
 
After optimization, we extract the posterior mean $\bar{f}_r$ of the latent spatial process
\begin{equation}
\bar{f}_r \;=\; \mathbb{E}_{\,q(f|\mathbf{s},\mathbf{p})}\!\left[f_r\right].
\end{equation}
 
\noindent and standardize it for use in the second GP. We use $q(\cdot)$ to refer to the variational posterior, $\mathbf{s}$ to refer to the $R \times 1$ vector of station counts for each unit, and $\mathbf{p}$ to refer to the corresponding vector of population for each unit.
 
\subsubsection{Predicting temperature with a second GP}
 
We use $\bar{f}_r$ to inform the covariance function $k_{2}(\cdot, \cdot)$ in the temperature model:
\begin{align}
y_r \mid \phi_r &\sim \mathcal{N}(\phi_r, \epsilon^2), \\
\phi_r &\sim \mathcal{GP}\!\big(\mu_r,\,k_2(r,r')\big), \\
\mu_r &= \boldsymbol{\beta}^\top \mathbf{w}_r + b.
\end{align}
where $\phi_r$ is an exact GP capturing residual spatial structure in temperature and $\epsilon$ is a small observation noise term learned by the model.
 
For the GP, we use a scaled Mat\'ern kernel with automatic relevance determination (ARD) acting on latitude, longitude, and the latent feature $\bar{f}_r$. ARD functions similarly to feature weighting: dimensions along which the function varies little are automatically assigned large length-scales, reducing their impact. A full explanation for kernel used and performance comparisons can be found in the appendix.
 
\subsection{Model fitting procedure}
 
Information on model fitting procedures is found in Appendix~\ref{sec:appendix_model_fit}.
 
\section{Results}
 
\subsection{Relationship between station density and income}
 
In this section, we demonstrate that WU data across all of North Carolina are preferentially sampled, and that this preference varies by county. To do so, we use a Poisson generalized linear mixed model (GLMM), where we allow the strength of the income effect to vary between counties.
 
Our observation model assumes that the number of personal weather stations in each region is Poisson distributed. This model is fit to areal data at the census-tract level. In the following, $s_r$ is the number of stations in census tract $r$, $p_r$ is the corresponding population and $x_r$ the corresponding income. We then define a Poisson GLMM over census tracts, where $\mu_r$ is the expected number of stations in tract $r$:
\begin{subequations}\label{eq:model}
    \begin{align}
        s_r &\sim \text{Poisson}(\mu_r), \label{eq:poisson-glmm}\\
        \log(\mu_r) &= \alpha_0 + (\alpha_1+\beta_{c(r)}) \log (x_r) + \alpha_2 \log(p_r),  \\
    \beta_{c(r)} &\sim \mathcal{N}(0, 10).
    \end{align}
\end{subequations}
In this observation model, $\alpha_0$ is the fixed intercept, $\alpha_1$ is the fixed effect for standardized log-income, and $\alpha_2$ is the fixed effect for log-population. $\beta_{c(r)}$ is the random slope for income associated with the county $c(r)$ to which unit $r$ belongs. Note that given the skewness of the income distribution, we log-transform and standardize the income dataset to have mean zero and variance one. Finally, we place a relatively weak prior on $\beta_{c(r)}$. The deviations are centered at zero (no county-specific bias assumed a priori).
 
After fitting this model on the July 2023 WU data, we find that the relationship between income and station density is highly significant: it obtains an equivalent p-value of less than $0.001$ as reported by PoissonBayesMixedGLM in Statsmodels \parencite{seabold2010statsmodels}. This result indicates that preferential sampling is present in the WU NC data that we use to fit our models.
 
To evaluate whether random income slopes improve model fit, we compare the Bayesian Information Criterion (BIC) of the GLMM to that of a fixed-effects GLM without county-specific income terms. The difference is substantial:
\begin{equation}
    \Delta \text{BIC} = \text{BIC}_\text{fixed} - \text{BIC}_\text{mixed} > 100.
\end{equation}
 
This result strongly favors the mixed-effects model. According to conventional interpretation \parencite{KassRaftery1995}, a difference in BIC greater than 10 provides very strong evidence in favor of the model with the lower BIC, so our result provides overwhelming support for including random effects. Thus, we conclude that the degree of preferential sampling varies significantly across North Carolina.
 
To visualize this spatial heterogeneity, we mapped the posterior mean income slope for each county, obtained from the GLMM. Specifically, for each county $c$ in North Carolina, we extracted the estimated total slope $\alpha_1 + \beta_c$. This quantity represents the strength of the association between income and station density within each county, after controlling for population.
 
\begin{figure}[htbp]
    \centering
    \includegraphics[width=0.65\textwidth]{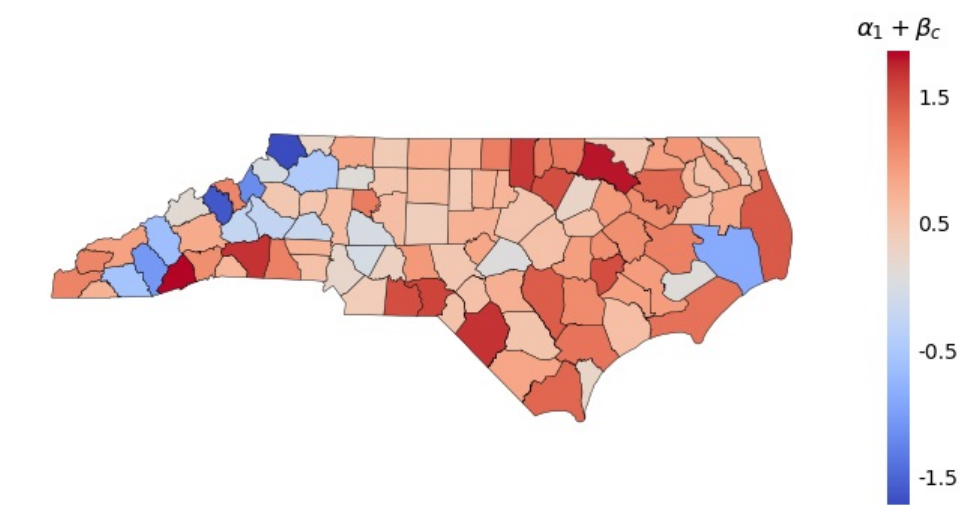}
    \caption{Map of income-to-sensor-intensity effects in NC counties based on averaged July 2023 data from WU.}
    \label{fig:income_effects_nc_map}
\end{figure}
 
The map in Figure~\ref{fig:income_effects_nc_map} demonstrates the extent to which preferential sampling varies across the state; some counties show strong socioeconomic bias, while others exhibit little to none.
 
\subsection{Model validation}
 
\begin{figure}[ht!]
    \centering
 
    \begin{subfigure}{0.95\linewidth}
        \centering
        \includegraphics[width=0.8\linewidth]{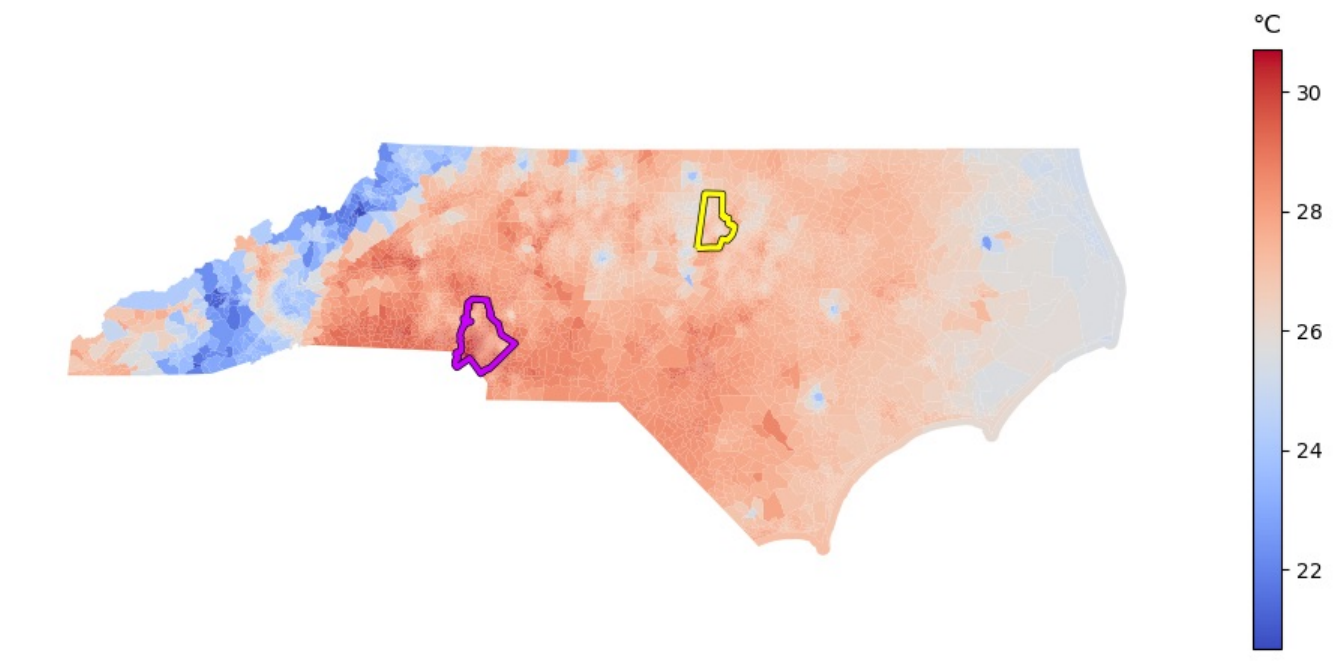}
        \caption{NC predictions on July 23, 2021 in the evening (7-8 P.M.)}
    \end{subfigure}
 
    \vspace{0.8em}
 
    \begin{subfigure}{0.4\linewidth}
        \centering
        \includegraphics[width=\linewidth]{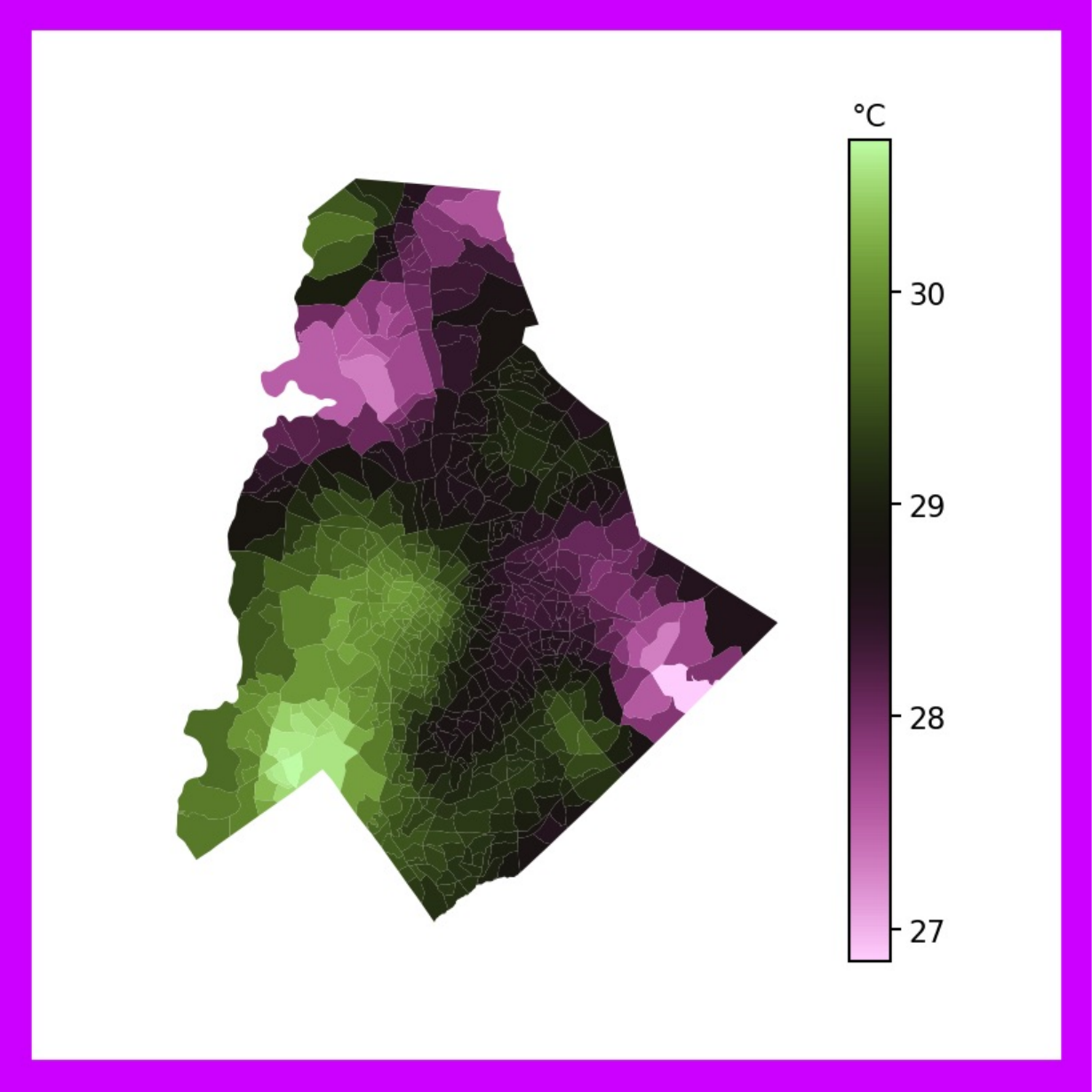}
        \caption{Charlotte portion of predictions}
    \end{subfigure}
    \hfill
    \begin{subfigure}{0.4\linewidth}
        \centering
        \includegraphics[width=\linewidth]{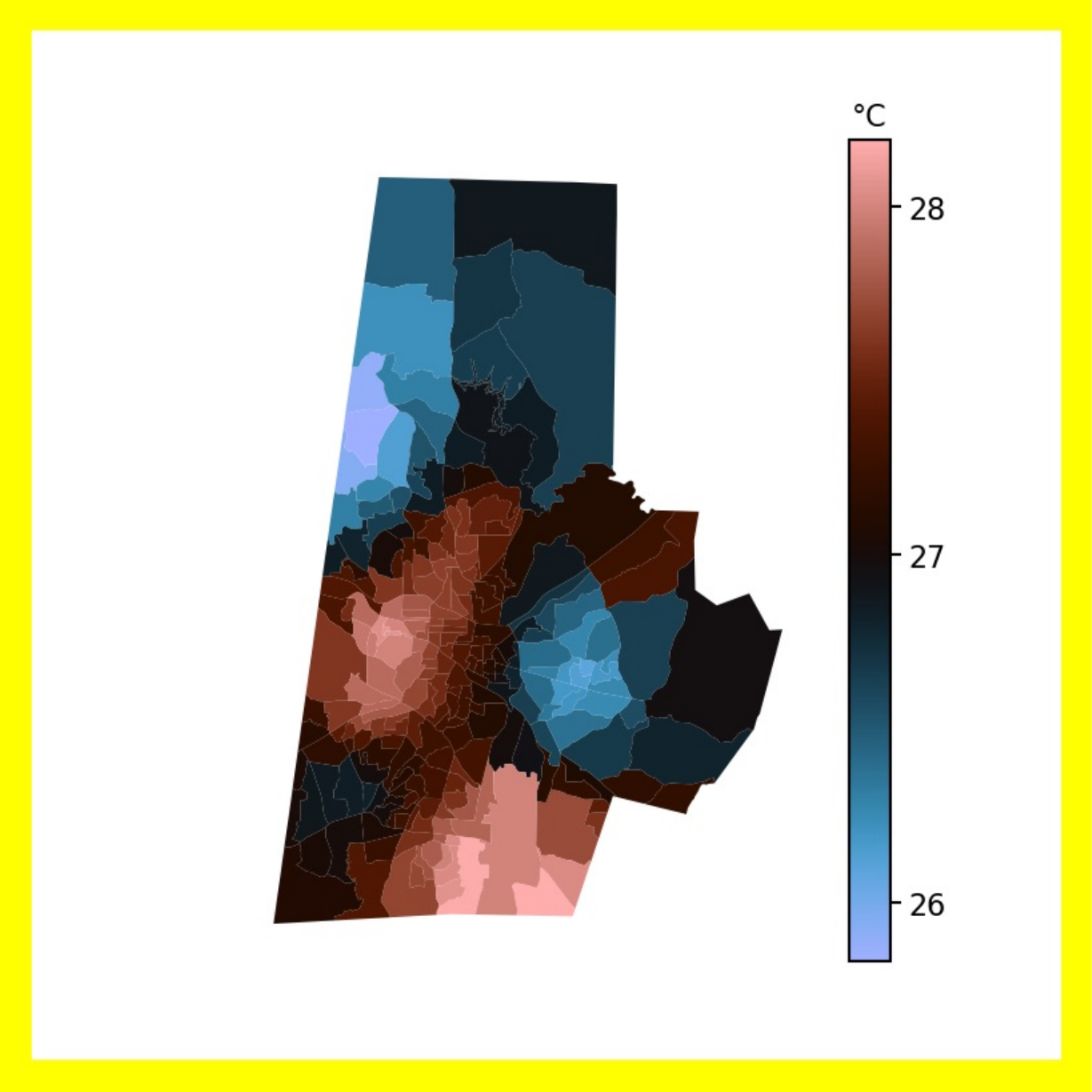}
        \caption{Durham portion of predictions}
    \end{subfigure}
 
    \caption{(Top) Visualization of full spatial map across North Carolina as obtained using LGCP+GP. (Bottom) Enlarged prediction maps of Durham and Charlotte obtained from the above NC map. These figures demonstrate the precise block-group-level detail.}
    \label{fig:nc-three-panels}
\end{figure}

In this section, we report the posterior mean predictions of surface temperature for four different models: (1) a base model using SVGP regression, (2) SVGPR+PS, (3) a base model using standard GP regression, and (4) LGCP+GP. The SVGP and standard GP models serve as baseline counterparts to SVGPR+PS and LGCP+GP respectively, allowing us to isolate the impact of adjusting for preferential sampling. These predictions are visualized as high-resolution maps, allowing for detailed comparisons across urban centers such as Durham and Raleigh (Figure~\ref{fig:nc-three-panels}).
 
To validate our predictions, we used high-resolution air temperature data from NOAA's UHI mapping campaign, coordinated by the CAPA Heat Watch program \parencite{NOAA}. The campaign provides raw mobile transect measurements and machine-learning--derived raster products combining traversal data with satellite imagery. We aggregated the NOAA raster data to census tracts and block groups by computing the mean raster value within each areal unit, yielding temperature estimates directly comparable to our model predictions, which are also produced at the areal level. Although the raw traversal data could be aggregated in a similar manner, this approach occasionally introduced extra noise due to uneven spatial coverage. We therefore use the aggregated raster product for validation.
 
For validation, we compared the aggregated NOAA values to predictions from all four models. At the census-tract and block-group levels, we computed the Pearson correlation between the model predictions and the corresponding NOAA value. These correlations measure each model's agreement with an external, non-preferentially-sampled dataset.
 
Additionally, we compare our predictions to ERA5-Land to determine if our models better match the ground truth (NOAA data) when it comes to fine-scale predictions. ERA5-Land is a global reanalysis product developed by the European Centre for Medium-Range Weather Forecasts (ECMWF), providing hourly surface variables at approximately 9-km resolution \parencite{essd-13-4349-2021}. While ERA5-Land performs well for statewide temperature patterns, its coarse resolution limits its ability to resolve neighborhood-level variability. Our intention in comparing our predictions to those of ERA5-Land is to demonstrate the advantages of correcting for preferential sampling as well as our ability to capture fine-scale variability.
 
\subsubsection{Correlation Results}
 
Table~\ref{tab:time_slot_svgpr_ps_summary} compares an SVGP baseline model to SVGPR+PS. It provides results for the three time slots from the NOAA UHI campaign, averaged across all cities included in the campaign. Due to its 9-km spatial resolution, ERA5-Land was upsampled via bilinear interpolation with resolution $0.05$ before being aggregated to areal units. The uncertainties for correlation values are also included after the $\pm$ symbol, and were calculated using the same method as \textcite{calhoun2024refining}, where we use effective sample size to obtain a more accurate uncertainty value. Results for individual cities and specific census group resolutions are available in the appendix. All models use 800 inducing points.
 
\begin{table}[H]
    \centering
    \small
    \setlength{\tabcolsep}{7pt}
    \begin{tabular}{l|c|c|c|c|c}
    Time Slot & ERA5-Land ($r_1$) & SVGPR ($r_2$) & SVGPR+PS ($r_3$) & $r_3 - r_1$ & $r_3 - r_2$ \\
    \hline
    Morning   & $-0.12 \pm 0.04$ & $0.44 \pm 0.04$ & $0.27 \pm 0.04$ & $0.39$ & $-0.17$ \\
    Afternoon & $0.25 \pm 0.05$  & $0.26 \pm 0.04$ & $0.43 \pm 0.04$ & $0.18$ & $0.17$ \\
    Evening   & $0.35 \pm 0.03$  & $0.44 \pm 0.03$ & $0.40 \pm 0.04$ & $0.05$ & $-0.04$ \\
    \end{tabular}
    \caption{Average Pearson correlation between model predictions and NOAA raster data, grouped by time slot and averaged across all cities and census group resolutions for that time slot. The last two columns report average improvement of SVGPR+PS over ERA5-Land and the baseline SVGPR model, respectively.}
    \label{tab:time_slot_svgpr_ps_summary}
\end{table}
 
Table~\ref{tab:lgcp_gp_correlations_averaged} compares model predictions for a standard GP with our LGCP+GP model. Again, this table provides results for all three time slots, averaged across all cities included in the NOAA UHI campaign. Results for individual cities and specific census group resolutions are available in the appendix. All models use 800 inducing points.
 
\begin{table}[H]
    \centering
    \small
    \setlength{\tabcolsep}{7pt}
    \begin{tabular}{l|c|c|c|c|c}
    Time Slot & ERA5-Land ($r_1$) & GP ($r_2$) & LGCP+GP ($r_3$) & $r_3 - r_1$ & $r_3 - r_2$ \\
    \hline
    Morning  & $-0.12 \pm 0.04$ & $0.47 \pm 0.04$ & $0.44 \pm 0.04$ & $0.56$ & $-0.04$ \\
    Afternoon & $0.25 \pm 0.04$  & $0.29 \pm 0.03$ & $0.31 \pm 0.03$ & $0.06$ & $0.02$  \\
    Evening  & $0.35 \pm 0.03$  & $0.48 \pm 0.03$ & $0.48 \pm 0.03$ & $0.13$ & $0.00$  \\
    \end{tabular}
    \caption{Average Pearson correlation between model predictions and NOAA raster data, grouped by time slot and averaged across all cities and census group resolutions for that time slot. The last two columns report average improvement of LGCP+GP over ERA5-Land and the baseline GP model, respectively.}
    \label{tab:lgcp_gp_correlations_averaged}
\end{table}
 
\subsubsection{Individual city results}
 
To further validate our results, we restrict our attention to Durham County. If we compare our SVGPR+PS predictions with those of the baseline SVGP, we may qualitatively assess where predictions change as a result of the correction and by how much. This comparison demonstrates the model's ability to adjust for preferential sampling and better recover the spatial field across undersampled regions.
 
\begin{figure}[H]
    \centering
    \resizebox{0.77\textwidth}{!}{%
        \begin{minipage}{\textwidth}
 
    \begin{subfigure}[t]{0.44\textwidth}
        \includegraphics[width=\linewidth]{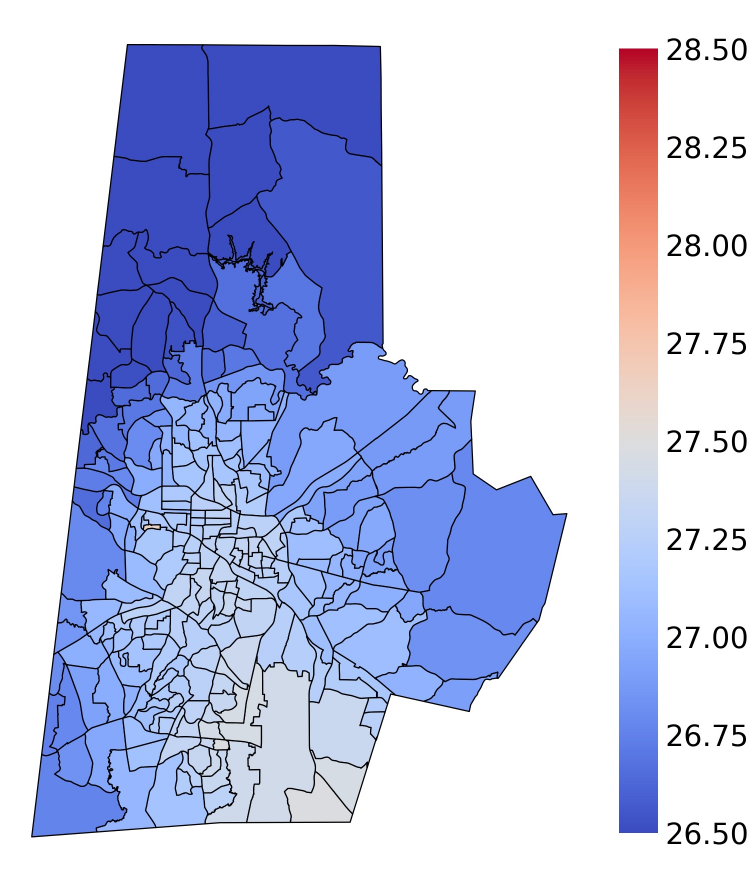}
        \caption{SVGPR}
    \end{subfigure}
    \hfill
    \begin{subfigure}[t]{0.44\textwidth}
        \includegraphics[width=\linewidth]{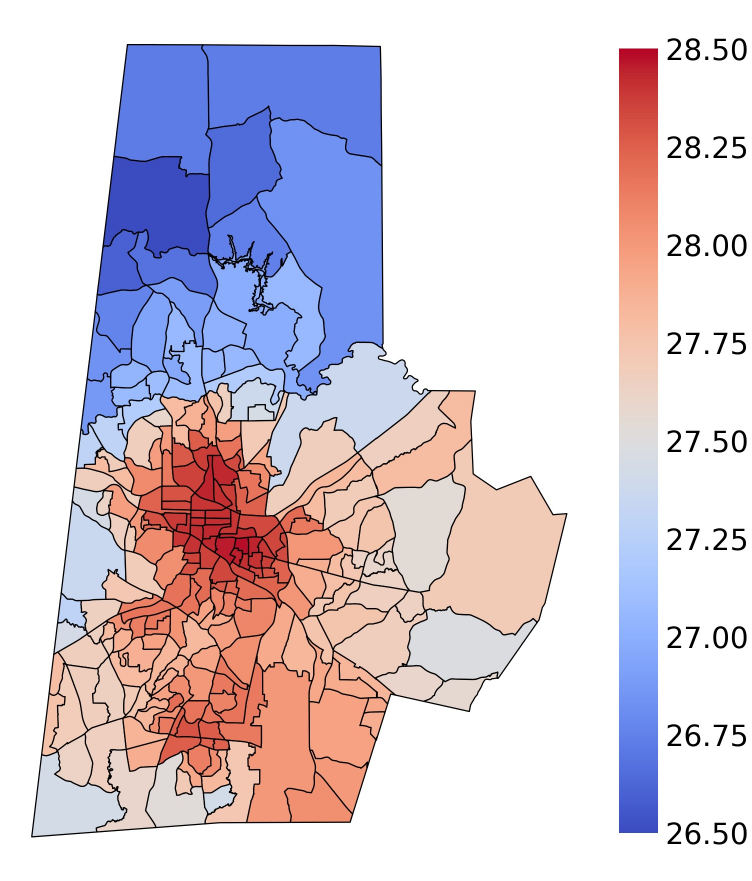}
        \caption{SVGPR+PS}
    \end{subfigure}
 
    \vspace{0.5em}
 
    \begin{subfigure}[t]{0.44\textwidth}
        \includegraphics[width=\linewidth]{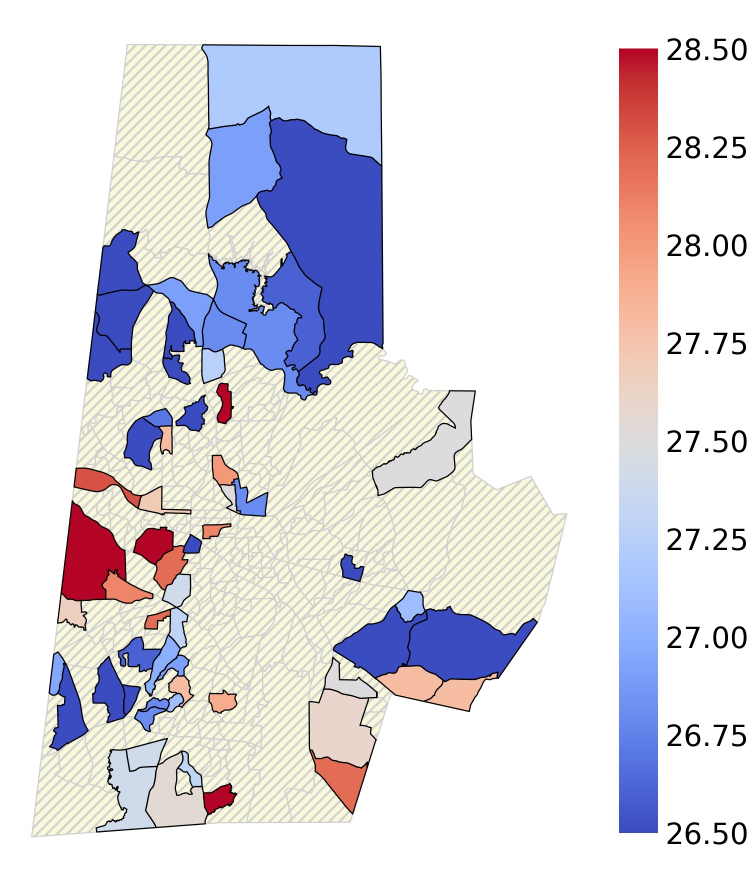}
        \caption{Observations (unobserved in striped beige)}
    \end{subfigure}
    \hfill
    \begin{subfigure}[t]{0.44\textwidth}
        \includegraphics[width=\linewidth]{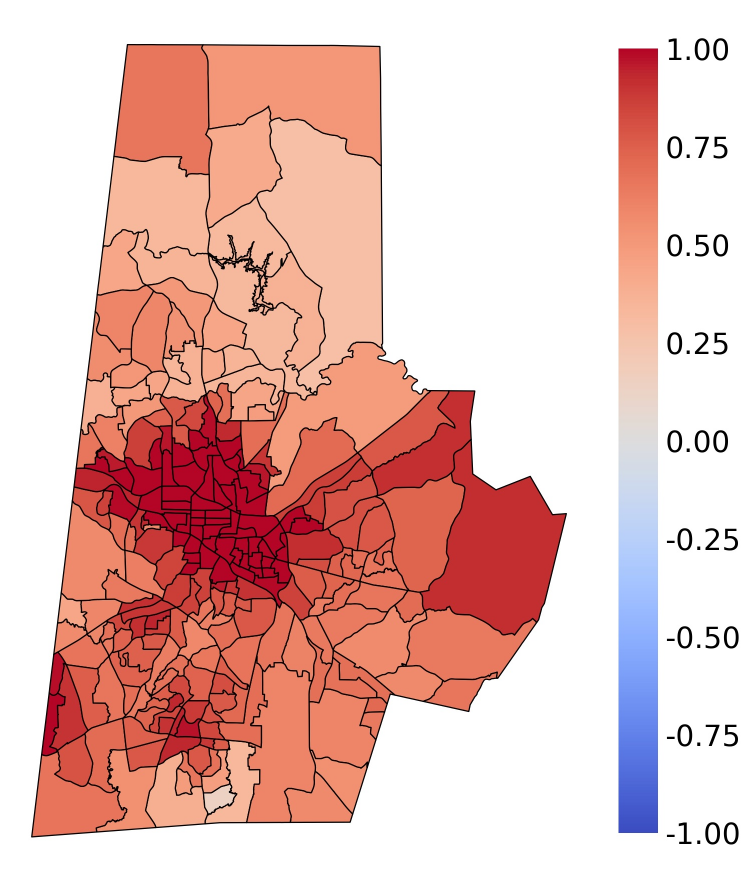}
        \caption{SVGPR+PS -- SVGPR}
    \end{subfigure}
 
    \end{minipage}
    }
 
    \caption{Evening (7-8 P.M.) temperatures on July 23, 2021 in Durham County, by census block group. All units are in degrees Celsius. (a) Uncorrected SVGPR model: mean estimate is 27.07 \degree C with standard deviation 0.28, (b) Corrected SVGPR+PS model: mean estimate is 27.86 \degree C with standard deviation 0.45, (c) Observed temperatures from WU: mean is 27.15 \degree C with standard deviation 0.88, (d) Difference between the corrected and uncorrected data: mean is 0.79 \degree C with standard deviation 0.24.}
    \label{fig:durham-noaa-comparison}
\end{figure}

As Figure~\ref{fig:durham-noaa-comparison} demonstrates, the SVGPR+PS model substantially improves correlation with the NOAA dataset when compared to standard SVGPR. This indicates that the preferential sampling adjustment improves model accuracy. For census tract data, we observed an improvement in correlation from $r=0.09$ to $r=0.49$ for a total difference of $0.40$ for the evening NOAA data. For block groups, we observed an improvement in correlation from $r=0.34$ to $r=0.42$ for a total difference of $0.08$ for the evening NOAA data.
 
Across cities and time slots, we find that both SVGPR+PS and LGCP+GP consistently achieve higher correlations with NOAA observations than the upsampled ERA5-Land. These gains are especially pronounced in areas with sparse sensor coverage or strong intra-urban temperature gradients, where flexible GP models better adapt to local structure. This indicates that our approaches capture fine-scale urban temperature variability more effectively than ERA5-Land, even after spatial upsampling.
 
However, while SVGPR+PS demonstrates clear improvements in cities such as Durham and Raleigh, it tends to fail in Charlotte (see appendix) where the estimated correlation drops sharply. By contrast, LGCP+GP yields stable results across all cities and spatial scales, even where SVGPR+PS fails. However, these improvements are mixed depending on the time of day. We see less of an advantage from the preferential sampling adjustment in the morning than in the afternoon, for instance (see Appendix~\ref{sec:appendix_individual_correlation_results} for details).
 
\subsection{Sensor network size dictates model predictive ability}
 
\begin{figure}[htbp]
    \centering
    \begin{subfigure}[t]{0.37\textwidth}
        \centering
        \includegraphics[width=\textwidth]{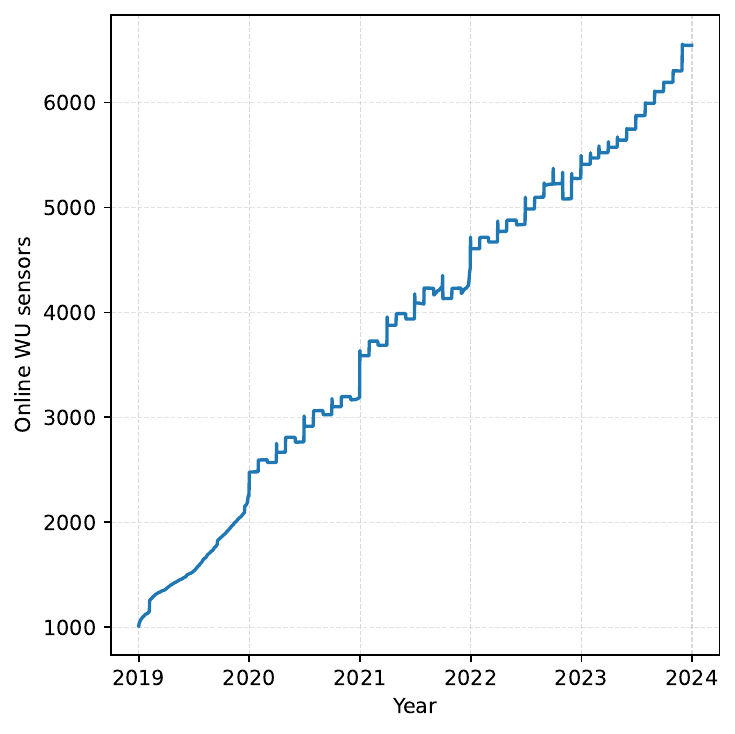}
        \caption{Growth of active WU sensors in NC from 2019 to 2024.}
        \label{fig:wu_network_over_time}
    \end{subfigure}
    \hfill
    \begin{subfigure}[t]{0.61\textwidth}
        \centering
        \includegraphics[width=\textwidth]{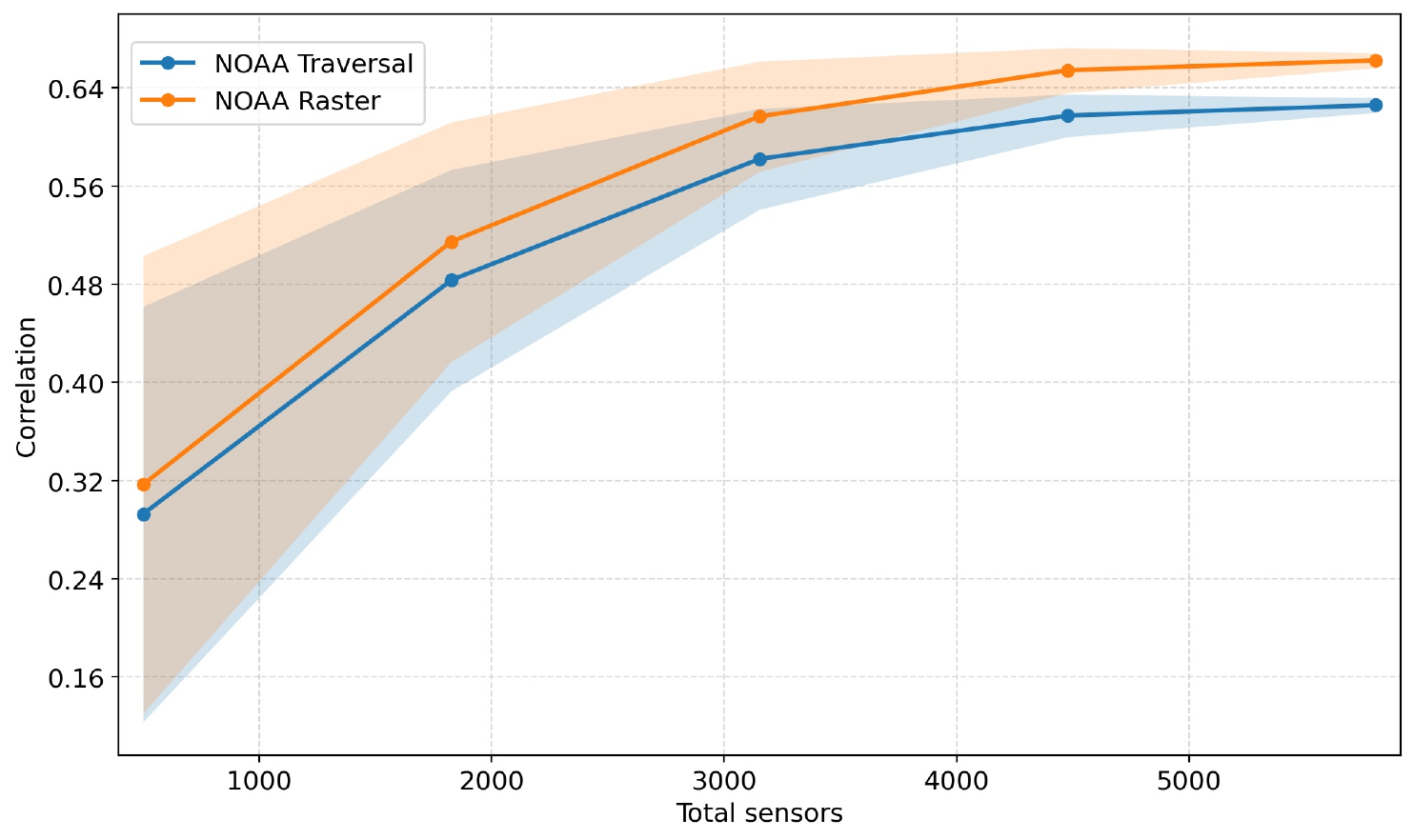}
        \caption{Predictive correlation with NOAA data as a function of sensor count for Charlotte (morning, 6--7 A.M.) block group data.}
        \label{fig:charlotte_compare_num_sensors}
    \end{subfigure}
    \caption{Expansion of the WU sensor network and the impact of its expansion on model predictive performance in Charlotte. As the number of available sensors increases over time, agreement between model predictions and NOAA UHI datasets improves. Aggregated traversal data is provided as an additional data point.}
    \label{fig:sensor_growth_and_performance}
\end{figure}
 
As the WU network has expanded over time, our ability to predict urban heat has improved. A denser sensor network provides greater spatial coverage, reducing interpolation error and limiting extrapolation from sparsely sampled areas. Figure~\ref{fig:wu_network_over_time} shows the steady growth of the WU network in NC from 2019 to 2024.
 
To quantify the impact of sensor density on predictive accuracy, we conducted a controlled subsampling analysis using the LGCP+GP model. Beginning with 500 sensors and increasing to the full network, we randomly selected subsets of stations from the training data, retrained the model using only those observations, and compared predictions to aggregated NOAA traversal and raster data using Pearson correlation. Each subsampling level was repeated across ten trials to reduce randomness, and we report the average NOAA correlation in Figure~\ref{fig:charlotte_compare_num_sensors}. This design isolates the effect of sensor density on model performance.
 
As the number of WU sensors increases, agreement between model predictions and the NOAA validation products improves steadily (Charlotte is shown in Figure~\ref{fig:charlotte_compare_num_sensors}; other cities in the appendix). This highlights the importance of citizen sensor networks for resolving fine-scale urban heat patterns: as sensor density grows, so does predictive accuracy.
 
This comes with caveats: while correlations increase most rapidly at low-to-moderate sensor counts, improvements taper as the final sensors are added. This suggests that additional sensors in already well-sampled areas may add limited new information while increasing redundancy. This is intuitive given that, in the case of GPs, predictive variance is reduced most efficiently by observations placed in regions of high prior uncertainty. Adding sensors in already dense areas primarily reduces local noise variance but does not substantially improve global interpolation accuracy, leading to diminishing gains in correlation. Hence, it is optimal for new sensors to be placed in a manner that expands spatial coverage, rather than adding to already well-documented areas.
 
\section{The importance of high-resolution data for epidemiology}
 
Our high-resolution data fill a longstanding gap in exposure assessment for urban health research. By aligning local heat conditions with socioeconomic and structural factors, researchers can study not just where extreme heat occurs, but also who is most affected and under what circumstances. As cities continue to experience more frequent and intense heat events, having heat data with high temporal and spatial resolution is increasingly important. These data will allow us to connect heat to health outcomes with more precision than ever before.
 
This work has important implications for epidemiology, particularly for understanding and mitigating the health impacts of urban heat exposure. In particular, a growing body of research highlights the importance of elevated nighttime temperatures, which may impair sleep quality, increase risk of mortality and hospitalizations, and exacerbate chronic conditions such as cardiovascular disease \parencite{Minor}. Nighttime heat may be especially consequential because individuals are more likely to be at home and therefore continuously exposed to local ambient conditions. At the same time, high-resolution daytime estimates are also critical for assessing risks faced by outdoor workers, unhoused populations, and other groups with substantial daytime exposure. Our high-resolution data provide an avenue for linking variation in both daytime and nighttime urban heat with clinical datasets, public health records, survey data, and demographic information. For example, hospital admissions for heat-related illnesses, emergency service calls during heatwaves, or wearable-device sleep monitoring data could all be paired with our temperature predictions to more precisely quantify the causal relationship between local heat exposure and adverse health outcomes. This would be a critical advance for environmental epidemiology, where exposure misclassification due to poor spatial coverage has historically limited risk estimation.
 
Beyond clinical outcomes, these high-resolution data products reveal how socioeconomic and structural inequities drive disparate levels of heat exposure. Localized differences in income, housing quality, tree cover, and access to cooling influence both baseline exposure and vulnerability to heat stress. With easy access to hourly bias-corrected data, researchers can identify where exposure is highest and assess how it interacts with social determinants of health. This makes it possible to identify unequal heat burdens across communities and better understand which neighborhoods face both elevated exposure and limited adaptive capacity.
 
These data products also support simulation studies of mitigation strategies such as urban greening, increased tree canopy, reflective roofing, or cool pavement infrastructure. Researchers can predict how hypothetical changes in land cover or surface properties would alter daytime and nighttime heat exposure at the neighborhood scale. For example, predicted reductions in nighttime minimum temperatures following tree canopy expansion could be translated into projected reductions in sleep disruption, cardiovascular strain, or heat-related hospitalizations using established exposure--response relationships. Similarly, researchers can assess how cooling interventions may differentially benefit outdoor workers or residents of high-density urban areas. By providing bias-corrected high-resolution temperature fields, our approach enables mitigation strategies to be evaluated not only for overall cooling effectiveness, but also for their potential to reduce disparities in heat exposure.
 
\section{Discussion}
 
Obtained results support our belief that the preferential sampling adjustment works as intended but only when necessary, and with certain caveats. Firstly, the preferential sampling adjustment is not expected to be equally necessary in all contexts. When sampling density is only weakly related to the latent spatial process, the corrected model converges to predictions similar to those of the uncorrected model. In other words, it reasonably learns the adjustment is unnecessary. Secondly, our ability to assess improvement is limited by the form of validation used, as explained in Appendix~\ref{sec:appendix_noaa_validation}. Thirdly, meteorological conditions such as wind speed and direction can disrupt expected urban heat patterns, particularly in the afternoon and evening. Future versions of the model would be improved by incorporating covariates like wind.
 
Although correcting for preferential sampling generally improves temperature prediction, this is not a substitute for incorporating strong covariates or collecting higher-quality observations. In this study, we focused primarily on income as a driver of sampling bias, but urban heat variability is also shaped by factors such as land use, building density, vegetation cover, and surface albedo, which were not explicitly modeled. Future work should integrate richer covariate information and replicate these analyses in other settings with independent validation data to further assess the robustness of preferential sampling adjustments.
 
Additionally, the use of fewer ($M = 800$) inducing points for this study was motivated by the fact that more inducing points performed minimally better. Furthermore, fewer inducing points contributes to a faster model. If increased to a much higher number of inducing points, the models would not have been able to run in such a short time.
 
While this study proposed two scalable frameworks to adjust for preferential sampling in citizen data, our analysis was restricted to single-hour intervals on specific days. A natural next step is to extend this framework to a fully spatiotemporal model capturing variation across space and time. For example, preliminary work by \textcite{calhoun_2025_spatiotemporal} demonstrates that spatiotemporal Gaussian processes with Student's~t likelihoods can model several months of noisy, crowdsourced data while maintaining calibrated uncertainty estimates. Integrating our preferential-sampling correction into such a spatiotemporal GP would enable hourly urban heat reanalysis across months to years while accounting for nonrandom sensor placement. This would also allow for the propagation of uncertainty over time, revealing how sampling bias evolves seasonally. Future work could explore natural-gradient or amortized variational inference for faster convergence, or deep and multi-output GP formulations to jointly model other meteorological variables such as humidity and radiation.
 
This analysis also suggests several directions for future work. First, we show that the degree of preferential sampling with respect to income varies by county. This suggests that the strength of the preferential-sampling adjustment itself should also vary spatially; our models could be extended to allow location-specific adjustment strength. Second, the LGCP+GP formulation is a two-stage estimator. To obviate the need for two steps, an alternative formulation could learn these separate GPs jointly, in a deep Gaussian process formulation \parencite{damianou2013deepgaussianprocesses}. However, such an approach would require updating the loss function to incorporate a non-Gaussian loss from the hidden layer (i.e., the Poisson likelihood). We leave these methods for future work.
 
\section{Conclusion}
 
We introduced two scalable Gaussian process frameworks that correct for preferential sampling in crowdsourced weather station data, enabling high-resolution, hourly temperature mapping across all of North Carolina. Both models substantially outperform ERA5-Land at the neighborhood scale and demonstrate that accounting for non-random sensor placement improves agreement with independent NOAA urban heat observations. As citizen sensor networks continue to grow, these bias-corrected temperature fields offer a practical foundation for epidemiological research linking heat exposure to health outcomes, and evaluation of urban heat mitigation strategies.

\clearpage
\appendix

\section{Model fitting procedure} \label{sec:appendix_model_fit}

All models are implemented in PyTorch \parencite{paszke2019pytorch}, using GPyTorch \parencite{gardner2018gpytorch} for Gaussian process primitives and Pyro \parencite{bingham2019pyro} for stochastic variational inference. For each dataset, inducing points are initialized from the centroids in $\mathcal{C}$. In the SVGPR+PS model, we use $M = 800$ inducing points initialized via $k$–means clustering and fix their locations during training. We chose this number of inducing points for both models because it provided a good trade-off between speed and accuracy: with $M=800$, model fitting typically required $2$–$3$ minutes per dataset, whereas $M=2000$ could take up to $15$ minutes. Although larger $M$ values usually yielded slightly lower MAE, we observed little improvement beyond $M=800$.

The SVGPR+PS model is trained with the Adam optimizer (learning rate $0.01$) and a cosine-annealing learning-rate schedule for 10 epochs with a batch size of 512. Within GPyTorch, the variational distribution over inducing values is parameterized via a full-rank Cholesky factor, allowing a dense covariance structure while maintaining numerical stability. Pyro is used to compute the evidence lower bound (ELBO), and model parameters such as kernel hyperparameters, variational mean and covariance, and observation-noise variance are learned by maximizing the ELBO.

The LGCP+GP model is fit in two stages. The LGCP uses $M = 800$ inducing points selected by a minimum-separation heuristic over centroids, with their locations optimized during training. The latent spatial field is inferred using stochastic variational inference in Pyro, with an SVI routine using $32$ Monte Carlo particles per iteration, and the Adam optimizer with learning rate $0.01$ for 300 iterations. In the second GP, an exact Gaussian process implemented in GPyTorch is trained using the posterior mean of the latent field from Stage~1 together with elevation and LST as predictors. The marginal log-likelihood is maximized with Adam with learning rate $0.1$ for 100 iterations.

\section{Background: sparse variational Gaussian processes}

Gaussian processes (GPs) provide a flexible nonparametric framework for modeling spatial structure, but their computational cost grows cubically in the number of observations due to the need to invert the covariance matrix. This $\mathcal{O}(N^3)$ scaling makes standard GP regression infeasible for statewide temperature modeling.

To overcome this limitation, we use a sparse variational Gaussian process (SVGP) rather than an exact GP. The SVGP introduces a set of $M\ll N$ inducing locations whose latent function values provide a compressed representation of the underlying GP. A variational distribution over these inducing values is learned, reducing computational complexity to \(\mathcal{O}(NM^2)\) \parencite{titsias_2009}.

The inducing locations, their variational distribution, and the kernel hyperparameters are optimized jointly by maximizing the evidence lower bound (ELBO). This objective fits the data through the expected log-likelihood term while regularizing the inducing-point approximation via a Kullback--Leibler penalty \parencite{titsias_2009}. The result is a GP approximation suitable for statewide, high-resolution temperature modeling.

\section{Background: shared latent process model}

In the original SLP \parencite{Diggle}, $f$ is the latent spatial process evaluated at locations $\{ \mathbf{x}_i\}_{i=1} ^n$ where each $\mathbf{x}_i \in \mathcal{D} \subset \mathbb{R}^2$. The goal of geostatistical inference is to predict unknown values of this process using the observations $\mathbf{X} = \{\mathbf{x}_1, ..., \mathbf{x}_n \}$.

In the following model, $\mu, \alpha, \beta$ and $\tau^2$ are scalars and $\mathbf{y}$ is the vector of $n$ observations obtained at the  locations in $\mathbf{X}$. The sampled locations $\mathbf{X}$ are modeled as an inhomogeneous Poisson process, and spatial process $f$ as a Gaussian process (GP) with zero mean and stationary covariance function $k_{\boldsymbol{\theta}}(\mathbf{x}, \mathbf{x'})$, where $\boldsymbol{\theta}$ are the kernel parameters. $\lambda(\mathbf{x})$ is the sampling intensity at location $\mathbf{x}$, which is directly dependent on the spatial process $f$. This relationship between $\lambda$ and $f$ accounts for preferential sampling; previous models assumed they were independent, an assumption that often does not hold true in geostatistical settings \parencite{Diggle}.
\begin{align}
  \mathbf{y}\mid \mathbf{X}, f &\sim \mathcal{N}_n(\mu \mathbf{1}_n + f(\mathbf{X}), \tau^2 I_n), \\
  f &\sim \mathcal{GP}(\mathbf{0}, k_{\boldsymbol{\theta}}(\cdot, \cdot)), \\
  \mathbf{X}\mid f &\sim \text{Poisson} (\lambda ( \mathbf{x} )), \\
  \lambda(\mathbf{x}) &= \text{exp}(\alpha + \beta f(\mathbf{x})).
\end{align}

\section{Log-likelihood derivation for SVGPR+PS}
\label{sec:appendix_svgpr_ps_log_likelihood}

We now give an expression for the log-likelihood. In our implementation, we simply optimize a reduced version of the log-likelihood with constants removed, as they do not affect the optimization. However, for the sake of rigorousness, we will provide a full derivation of the log-likelihood. Starting from the Poisson pmf, we have:
\begin{align}
    p(s_r \mid \lambda_r)
= \frac{(\lambda_r A_r)^{s_r}\exp(-\lambda_r A_r)}{s_r!}.
\end{align}

\noindent The expression for log-likelihood for a single region is then:
\begin{align}
    \log p(s_r \mid \lambda_r)
= s_r \log \lambda_r + s_r \log A_r - \lambda_r A_r - \log(s_r!).
\end{align}

\noindent Assuming conditional independence across regions given the intensity field:
\begin{align}
    \log p(\{s_r\}_{r=1}^R \mid \{\lambda_r\}_{r=1}^R)
= \sum_{r=1}^R \Big[ s_r \log \lambda_r - \lambda_r A_r \Big]
+ \sum_{r=1}^R \Big[ s_r \log A_r - \log(s_r!) \Big].
\end{align}

The second sum does not depend on $(\alpha_0,\alpha_1,f)$ since $A_r$ and $s_r$ are fixed data, so in implementation we just optimize the simplified objective:
\begin{align}
    \log p(\{s_r\}_{r=1}^R \mid \{\lambda_r\}_{r=1}^R)
\;\propto\;
\sum_{r=1}^R \Big[ s_r \log \lambda_r - \lambda_r A_r \Big].
\end{align}

\section{LGCP+GP equivalence of the preferential sampling adjustment}
\label{sec:appendix_lgcp_gp_equivalence}

In this section, we provide further details motivating the use of the LGCP+GP formulation as a preferential sampling correction. First, we show that, given the true sampling intensity, we can recover the exact Gaussian process.

Recall the standard preferential sampling definition from \textcite{Diggle}:
\begin{align}
  \mathbf{y}\mid \mathbf{X}, f &\sim \mathcal{N}_n(\mu \mathbf{1}_n + f(\mathbf{X}), \tau^2 I_n), \\
  f &\sim \mathcal{GP}(\mathbf{0}, k_{\boldsymbol{\theta}}(\cdot, \cdot)), \\
  \mathbf{X}\mid f &\sim \text{Poisson} (\lambda ( \mathbf{x} )), \\
  \lambda(\mathbf{x}) &= \text{exp}(\alpha + \beta f(\mathbf{x})).
\end{align}
While somewhat trivial to show, in the case where we know $\lambda$, then $\log \lambda \propto f(\mathbf{x})$. Because $\mathbb{E}[y_i] = \mu + \mathbb{E}[f(\mathbf{x}_i)]$, then $\log \lambda_i \propto f(\mathbf{x}_i)\propto y_i$. Due to this relationship, we merely need to learn the multiplicative constant and offset (i.e., linear coefficient and bias) such that $\mathbb{E}[\log \lambda] = \gamma_0 +\gamma_1 \mathbb{E}[Y]$.

\subsubsection*{In the case of unknown intensity function}
Given that we rarely know the true intensity, we should consider the extent to which the sparse variational Gaussian process allows for unbiased estimation of the log sampling intensity, and whether this approach is preferable to alternative methods.

Previous techniques for preferential sampling have opted to omit learning this function directly, by instead approximating the intensity function using a kernel density estimator \cite{hsiao2025inverse}. This technique goes by Inverse Sampling Intensity Weighting. However, the motivation for such a two-stage estimator can be traced back to \textcite{reich_2012_AccountingDesignAnalysis}, who suggested that this technique is both computationally tractable and potentially allows for more flexibility, too. They provide the following solution for estimating the uncertainty of the learned field. That is, under the given model formulation, if we have an estimator $\hat{\lambda}$, such that:
\begin{equation}
    \log \lambda \sim \mathcal{N}(\log \hat{\lambda}, \Sigma_\lambda)
\end{equation}
Then the field estimate's error has uncertainty $\tau^2I + b^2\Sigma_\lambda$, where $b$ denotes the multiplicative constant of proportionality. The key innovation in our paper is then:
\begin{enumerate}
    \item We approximate the latent field using a sparse variational Gaussian process.
    \item Rather than merely incorporating this latent field into the mean function, we multiply our spatial kernel with a kernel for this latent field. This allows for the model to adjust its estimates on preferential sampling locally (i.e., over a city) rather than globally (i.e., over the whole state of North Carolina).
\end{enumerate}

Given this technique, it may be worth asking to what extent the sparse variational Gaussian process provides a useful estimate $\hat{\lambda}$. Estimating heterogeneous Poisson processes is a difficult task; exact inference is computationally expensive. To overcome this issue, a common technique is the kernel smoothing estimator. However, as \textcite{hsiao2025inverse} admit, ``nonparametric kernel smoothing estimators of the intensity [...] have few theoretical guarantees [citing \textcite{guan_2008_ConsistentNonparametricIntensity}].'' The key problem with kernel smoothing estimators is their consistency. Specifically, kernel smoothing estimators lack consistency guarantees under heterogeneous Poisson processes and do not explicitly model the latent spatial field driving preferential sampling. In contrast, approximating the intensity surface via a sparse variational Gaussian process preserves the structural dependence implied by the preferential sampling model while remaining computationally tractable.

In summary, the LGCP+GP formulation closely parallels the two-stage approach of \textcite{reich_2012_AccountingDesignAnalysis}: we first estimate the log sampling intensity surface and then incorporate it as a predictor in the response model to adjust for preferential sampling. The primary distinction lies in implementation: rather than using kernel density estimators or MCMC-based inference, we estimate the latent intensity via a variational log-Gaussian Cox process and embed it within a scalable Gaussian process framework. Thus, LGCP+GP can be viewed as a computationally efficient GP-based realization of the aforementioned strategy for preferential sampling correction.

\section{LGCP+GP kernel formulation}

We use a Matérn kernel with smoothing parameter $\nu = 1.5$ to allow moderately smooth spatial variation while retaining flexibility to capture localized gradients. This kernel is appropriate for temperature fields that vary smoothly across space but still exhibit spatial gradients due to land cover, urban structure, and topography.
\begin{align}
k_2(r,r') &= \sigma_1^2 \, k_{\text{Matérn}, 1}(r,r') + \sigma_2^2 \, k_{\text{Matérn}, 2}(r, r') \, k_{\text{Linear}}(r,r'), \\[6pt]
k_{\text{Matérn}, j}(r,r') &= \left( 1 + \frac{\sqrt{3}\,\|\mathbf{c}_r - \mathbf{c}_{r'}\|}{\ell_j} \right)  \exp\!\left(-\frac{\sqrt{3}\,\|\mathbf{c}_r - \mathbf{c}_{r'}\|}{\ell_j} \right), \quad j = 1, 2, \\[6pt]
k_{\text{Linear}}(r,r') &= \,\bar{f}_r \cdot \bar{f}_{r'}.
\end{align}

We include a linear kernel over the standardized latent field in addition to the baseline Matérn component to allow the strength of spatial covariance to scale with inferred sampling intensity. The first Matérn term captures residual spatial smoothness in temperature, while the second multiplicative term  increases covariance in regions where both the latent intensity and spatial proximity are high. Intuitively, if preferential sampling is stronger in certain areas, we allow temperature variation in those areas to be amplified.

We additionally experimented with adding the latent field as a covariate in the linear mean \parencite{reich_2012_AccountingDesignAnalysis}, which is a more customary method of integrating the latent. While this yielded similar predictive results, the above framework performed slightly better. Thus, we decided to include the latent field in the covariance structure.

\section{Individual correlation results for cities}
\label{sec:appendix_individual_correlation_results}

In this section, we provide non-averaged correlation results for each city. The following three tables compare the LGCP+GP model to a standard GP. The comparison is made against aggregated raster images from NOAA's Urban Heat Mapping campaign, averaged across census tracts and block groups.

\begin{table}[H]
\centering
\small
\setlength{\tabcolsep}{9pt}
\begin{tabular}{|l|l|c|c|c|c|}
\hline
City 
& Census Group
& \begin{tabular}{c} GP Corr \\ $r_1$ \end{tabular}
& \begin{tabular}{c} LGCP+GP Corr \\ $r_2$ \end{tabular}
& ERA5-Land
& $r_2 - r_1$ \\
\hline
Durham & Census Tract 
& $0.58 \pm 0.07$ 
& $0.56 \pm 0.08$ 
& -$0.05 \pm 0.12$ 
& -$0.02$ \\
Durham & Block Group 
& $0.55 \pm 0.09$ 
& $0.55 \pm 0.08$ 
& $0.05 \pm 0.12$ 
& -$0.01$ \\
Raleigh & Census Tract 
& $0.45 \pm 0.12$ 
& $0.46 \pm 0.12$ 
& -$0.15 \pm 0.09$ 
& $0.01$ \\
Raleigh & Block Group 
& $0.38 \pm 0.12$ 
& $0.39 \pm 0.12$ 
& -$0.18 \pm 0.09$ 
& $0.01$ \\
Asheville & Census Tract 
& $0.31 \pm 0.16$ 
& $0.20 \pm 0.17$ 
& -$0.05 \pm 0.11$ 
& -$0.11$ \\
Asheville & Block Group 
& $0.20 \pm 0.15$ 
& $0.19 \pm 0.15$ 
& -$0.06 \pm 0.10$ 
& -$0.01$ \\
Charlotte & Census Tract 
& $0.69 \pm 0.13$ 
& $0.53 \pm 0.13$ 
& -$0.26 \pm 0.14$ 
& -$0.16$ \\
Charlotte & Block Group 
& $0.62 \pm 0.12$ 
& $0.61 \pm 0.12$ 
& -$0.22 \pm 0.15$ 
& -$0.01$ \\
\hline
\multicolumn{2}{|c|}{\textbf{Averages}}
& $0.47 \pm 0.04$
& $0.44 \pm 0.04$
& -$0.12 \pm 0.04$
& -$0.04$ \\
\hline

\end{tabular}
\caption{Correlations of the standard GP and LGCP+GP model with aggregated NOAA raster data during the morning time slot (6--7 A.M.). Both models use 800 inducing points.}
\label{tab:correlation_analysis_morning_raster_only}
\end{table}

\begin{table}[H]
\centering
\small
\setlength{\tabcolsep}{9pt}
\begin{tabular}{|l|l|c|c|c|c|}
\hline
City 
& Census Group
& \begin{tabular}{c} GP Corr \\ $r_1$ \end{tabular}
& \begin{tabular}{c} LGCP+GP Corr \\ $r_2$ \end{tabular}
& ERA5-Land
& $r_2 - r_1$ \\
\hline
Durham & Census Tract 
& -$0.10 \pm 0.05$ 
& -$0.11 \pm 0.05$ 
& $0.30 \pm 0.06$ 
& $0.00$ \\
Durham & Block Group 
& -$0.27 \pm 0.05$ 
& -$0.06 \pm 0.07$ 
& $0.31 \pm 0.08$ 
& $0.21$ \\
Raleigh & Census Tract 
& $0.26 \pm 0.09$ 
& $0.20 \pm 0.08$ 
& $0.06 \pm 0.09$ 
& -$0.06$ \\
Raleigh & Block Group 
& $0.05 \pm 0.07$ 
& $0.02 \pm 0.08$ 
& $0.07 \pm 0.10$ 
& -$0.03$ \\
Asheville & Census Tract 
& $0.40 \pm 0.11$ 
& $0.42 \pm 0.11$ 
& $0.36 \pm 0.12$ 
& $0.02$ \\
Asheville & Block Group 
& $0.43 \pm 0.08$ 
& $0.44 \pm 0.07$ 
& $0.29 \pm 0.12$ 
& $0.01$ \\
Charlotte & Census Tract 
& $0.76 \pm 0.11$ 
& $0.76 \pm 0.10$ 
& $0.29 \pm 0.14$ 
& $0.00$ \\
Charlotte & Block Group 
& $0.78 \pm 0.10$ 
& $0.78 \pm 0.10$ 
& $0.30 \pm 0.15$ 
& $0.00$ \\
\hline
\multicolumn{2}{|c|}{\textbf{Averages}}
& $0.29 \pm 0.03$
& $0.31 \pm 0.03$
& $0.25 \pm 0.04$
& $0.02$ \\
\hline

\end{tabular}
\caption{Correlations of the standard GP and LGCP+GP model with aggregated NOAA raster data during the afternoon time slot (3--4 P.M.). Both models use 800 inducing points.}
\label{tab:correlation_analysis_afternoon_raster_only}
\end{table}

\begin{table}[H]
\centering
\small
\setlength{\tabcolsep}{9pt}
\begin{tabular}{|l|l|c|c|c|c|}
\hline
City 
& Census Group
& \begin{tabular}{c} GP Corr \\ $r_1$ \end{tabular}
& \begin{tabular}{c} LGCP+GP Corr \\ $r_2$ \end{tabular}
& ERA5-Land
& $r_2 - r_1$ \\
\hline

Durham & Census Tract 
& $0.33 \pm 0.08$ 
& $0.26 \pm 0.08$ 
& $0.36 \pm 0.07$ 
& -$0.06$ \\

Durham & Block Group 
& $0.49 \pm 0.09$ 
& $0.49 \pm 0.08$ 
& $0.39 \pm 0.09$ 
& -$0.01$ \\

Raleigh & Census Tract 
& $0.32 \pm 0.09$ 
& $0.34 \pm 0.08$ 
& $0.00 \pm 0.10$ 
& $0.02$ \\

Raleigh & Block Group 
& $0.34 \pm 0.10$ 
& $0.37 \pm 0.10$ 
& -$0.01 \pm 0.10$ 
& $0.03$ \\

Asheville & Census Tract 
& $0.45 \pm 0.10$ 
& $0.45 \pm 0.09$ 
& $0.29 \pm 0.09$ 
& $0.01$ \\

Asheville & Block Group 
& $0.40 \pm 0.09$ 
& $0.40 \pm 0.09$ 
& $0.36 \pm 0.07$ 
& $0.01$ \\

Charlotte & Census Tract 
& $0.77 \pm 0.08$ 
& $0.77 \pm 0.08$ 
& $0.68 \pm 0.10$ 
& $0.00$ \\

Charlotte & Block Group 
& $0.76 \pm 0.10$ 
& $0.76 \pm 0.09$ 
& $0.69 \pm 0.10$ 
& $0.00$ \\

\hline
\multicolumn{2}{|c|}{\textbf{Averages}}
& $0.48 \pm 0.03$
& $0.48 \pm 0.03$
& $0.35 \pm 0.03$
& $0.00$ \\
\hline

\end{tabular}
\caption{Correlations of the standard GP and LGCP+GP model with aggregated NOAA raster data during the evening time slot (7--8 P.M.). Both models use 800 inducing points.}
\label{tab:correlation_analysis_evening_raster_only}
\end{table}

Similarly, the following three tables compare the SVGPR+PS model to a standard SVGP regression model without preferential sampling adjustment. The comparison is made against aggregated raster images from NOAA's Urban Heat Mapping campaign, averaged across census tracts and block groups.

\begin{table}[H]
\centering
\resizebox{\textwidth}{!}{%
\begin{tabular}{|l|l|c|c|c|c|}
\hline
City
& Census Group
& \begin{tabular}{c} \rule{0pt}{16pt} SVGPR Corr \\ $r_1$ \end{tabular}
& \begin{tabular}{c} \rule{0pt}{16pt} SVGPR+PS Corr \\ $r_2$ \end{tabular}
& \begin{tabular}{c} \rule{0pt}{16pt} ERA5-Land \end{tabular}
& \begin{tabular}{c} $r_2 - r_1$ \end{tabular} \\
\hline
Durham     & Census Tract & $0.33 \pm 0.09$ & $0.35 \pm 0.08$ & -$0.05 \pm 0.12$ & $0.02$ \\
Durham     & Block Group  & $0.34 \pm 0.10$ & $0.20 \pm 0.12$ & $0.05 \pm 0.12$  & -$0.14$ \\
Raleigh    & Census Tract & $0.52 \pm 0.09$ & $0.25 \pm 0.09$ & -$0.15 \pm 0.09$ & -$0.27$ \\
Raleigh    & Block Group  & $0.38 \pm 0.11$ & $0.22 \pm 0.10$ & -$0.18 \pm 0.09$ & -$0.16$ \\
Asheville  & Census Tract & $0.44 \pm 0.14$ & $0.55 \pm 0.13$ & -$0.05 \pm 0.11$ & $0.12$ \\
Asheville  & Block Group  & $0.22 \pm 0.15$ & $0.49 \pm 0.09$ & -$0.06 \pm 0.10$ & $0.27$ \\
Charlotte  & Census Tract & $0.65 \pm 0.13$ & -$0.23 \pm 0.16$ & -$0.26 \pm 0.14$ & -$0.88$ \\
Charlotte  & Block Group  & $0.64 \pm 0.12$ & $0.29 \pm 0.14$ & -$0.22 \pm 0.15$ & -$0.35$ \\
\hline
\multicolumn{2}{|c|}{\textbf{Averages}}
& $0.44 \pm 0.04$
& $0.27 \pm 0.04$
& -$0.12 \pm 0.04$
& -$0.17$ \\
\hline
\end{tabular}
}
\caption{Correlations of the SVGPR and SVGPR+PS models with aggregated NOAA raster data during the morning time slot (6--7 A.M.). Both models use 800 inducing points.}
\label{tab:svgpr_ps_morning_raster}
\end{table}

\begin{table}[H]
\centering
\resizebox{\textwidth}{!}{%
\begin{tabular}{|l|l|c|c|c|c|}
\hline
City
& Census Group
& \begin{tabular}{c} \rule{0pt}{16pt} SVGPR Corr \\ $r_1$ \end{tabular}
& \begin{tabular}{c} \rule{0pt}{16pt} SVGPR+PS Corr \\ $r_2$ \end{tabular}
& \begin{tabular}{c} \rule{0pt}{16pt} ERA5-Land \end{tabular}
& \begin{tabular}{c} $r_2 - r_1$ \end{tabular} \\
\hline
Durham        & Census Tract & -$0.07 \pm 0.08$ & $0.41 \pm 0.07$ & $0.30 \pm 0.06$ & $0.49$ \\
Durham        & Block Group  & -$0.07 \pm 0.08$ & $0.30 \pm 0.12$ & $0.31 \pm 0.08$ & $0.37$ \\
Raleigh       & Census Tract & $0.32 \pm 0.09$  & $0.40 \pm 0.08$ & $0.06 \pm 0.09$ & $0.08$ \\
Raleigh       & Block Group  & $0.05 \pm 0.09$  & $0.45 \pm 0.09$ & $0.07 \pm 0.10$ & $0.40$ \\
Asheville     & Census Tract & $0.51 \pm 0.11$  & $0.58 \pm 0.12$ & $0.36 \pm 0.12$ & $0.07$ \\
Asheville     & Block Group  & $0.53 \pm 0.08$  & $0.53 \pm 0.10$ & $0.29 \pm 0.12$ & $0.00$ \\
Charlotte     & Census Tract & $0.38 \pm 0.15$  & $0.18 \pm 0.13$ & $0.29 \pm 0.14$ & -$0.20$ \\
Charlotte     & Block Group  & $0.39 \pm 0.16$  & $0.57 \pm 0.12$ & $0.30 \pm 0.15$ & $0.18$ \\
\hline
\multicolumn{2}{|c|}{\textbf{Averages}}
& $0.26 \pm 0.04$
& $0.43 \pm 0.04$
& $0.25 \pm 0.05$
& $0.17$ \\
\hline
\end{tabular}
}
\caption{Correlations of the SVGPR and SVGPR+PS models with aggregated NOAA raster data during the afternoon time slot (3--4 P.M.). Both models use 800 inducing points.}
\label{tab:svgpr_ps_afternoon_raster}
\end{table}

\begin{table}[H]
\centering
\resizebox{\textwidth}{!}{%
\begin{tabular}{|l|l|c|c|c|c|}
\hline
\text{City} 
& \text{Census Group}
& \begin{tabular}{c} \rule{0pt}{16pt} SVGPR Corr \\ $r_1$ \end{tabular}
& \begin{tabular}{c} \rule{0pt}{16pt} SVGPR+PS Corr \\ $r_2$ \end{tabular}
& \begin{tabular}{c} \rule{0pt}{16pt} ERA5-Land \end{tabular}
& \begin{tabular}{c} $r_2 - r_1$ \end{tabular} \\
\hline
Durham & Census Tract & $0.09 \pm 0.08$ & $0.49 \pm 0.06$ & $0.36 \pm 0.07$ & $0.40$ \\
Durham & Block Group & $0.34 \pm 0.09$ & $0.42 \pm 0.10$ & $0.39 \pm 0.09$ & $0.08$ \\
Raleigh & Census Tract & $0.30 \pm 0.09$ & $0.37 \pm 0.08$ & $0.00 \pm 0.10$ & $0.07$ \\
Raleigh & Block Group & $0.28 \pm 0.10$ & $0.42 \pm 0.07$ & -$0.01 \pm 0.10$ & $0.14$ \\
Asheville & Census Tract & $0.47 \pm 0.10$ & $0.64 \pm 0.10$ & $0.29 \pm 0.09$ & $0.17$ \\
Asheville & Block Group  & $0.54 \pm 0.08$ & $0.49 \pm 0.11$ & $0.36 \pm 0.07$ & -$0.05$ \\
Charlotte & Census Tract & $0.75 \pm 0.09$ & $0.20 \pm 0.15$ & $0.68 \pm 0.10$ & -$0.55$ \\
Charlotte & Block Group & $0.75 \pm 0.10$ & $0.13 \pm 0.18$ & $0.69 \pm 0.10$ & -$0.63$ \\
\hline
\multicolumn{2}{|c|}{\textbf{Averages}}
& $0.44 \pm 0.03$ 
& $0.40 \pm 0.04$
& $0.35 \pm 0.03$
& -$0.04$ \\
\hline
\end{tabular}
}
\caption{Correlations of the SVGPR and SVGPR+PS models with aggregated NOAA raster data during the evening time slot (7--8 P.M.). Both models use 800 inducing points.}
\label{tab:svgpr_ps_evening_raster}
\end{table}

\section{Sensor-to-correlation comparisons for other cities}
\label{sec:appendix_sensor_to_correlation}

 \begin{figure}[H]
    \centering
    \includegraphics[width=0.7\textwidth]{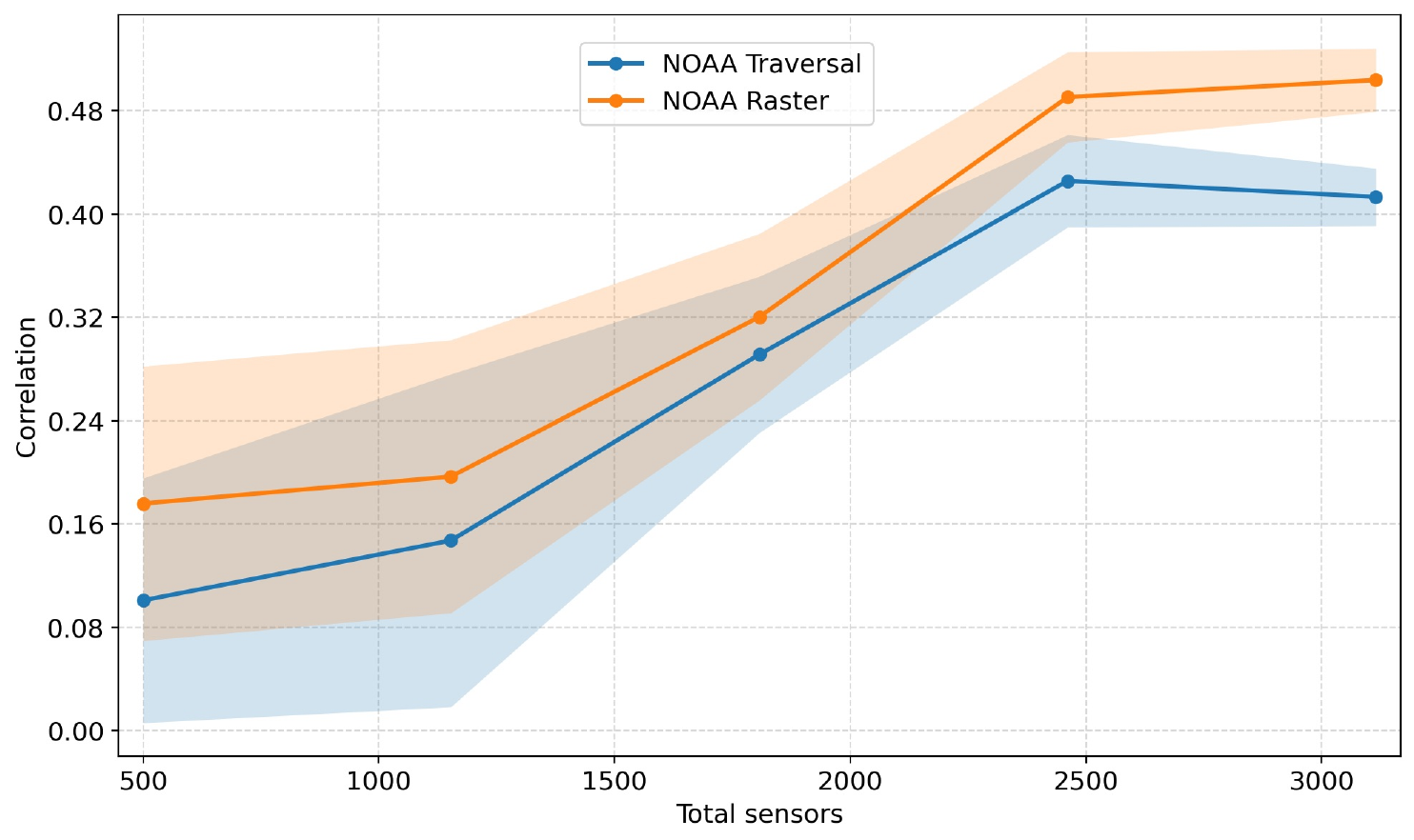}
    \caption{Predictive correlation with NOAA Urban Heat Island (UHI) campaign data as a function of total sensor count for Durham on July 23, 2021 (6–7 A.M.). Correlations are computed between model-predicted block-group temperatures (using randomly selected subsets of sensors from that day) and (i) NOAA mobile traversal measurements (blue) and (ii) NOAA raster products (orange), both aggregated to block groups. Shaded bands indicate 95\% bootstrap confidence intervals. Predictive agreement increases with sensor density, with consistently higher correlations against NOAA raster data.}
    \label{fig:durham_compare_num_sensors}
\end{figure}

 \begin{figure}[H]
    \centering
    \includegraphics[width=0.7\textwidth]{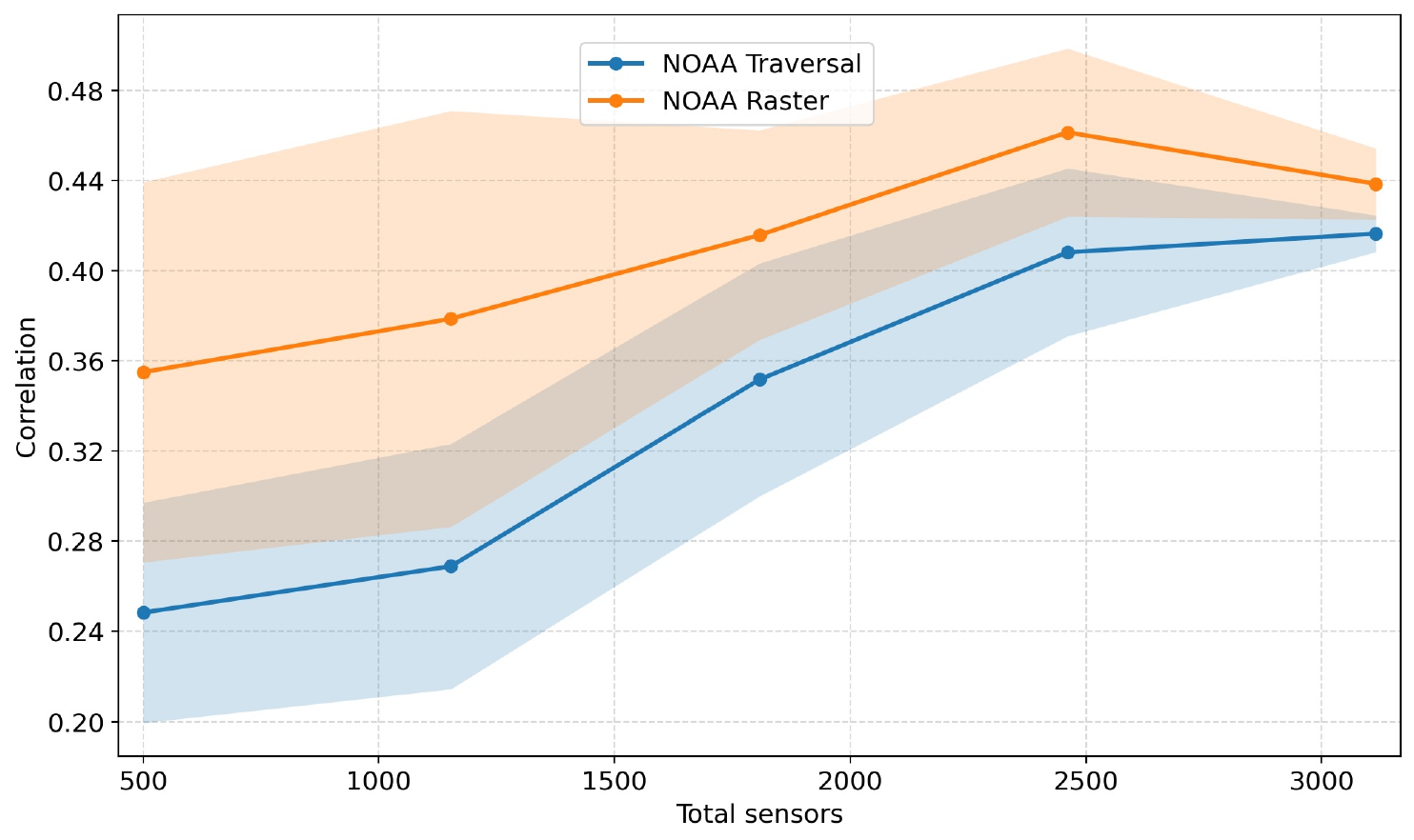}
    \caption{Predictive correlation with NOAA Urban Heat Island (UHI) campaign data as a function of total sensor count for Raleigh on July 23, 2021 (6–7 A.M.). Correlations are computed between model-predicted block-group temperatures (using randomly selected subsets of sensors from that day) and (i) NOAA mobile traversal measurements (blue) and (ii) NOAA raster products (orange), both aggregated to block groups. Shaded bands indicate 95\% bootstrap confidence intervals. Predictive agreement increases with sensor density, with consistently higher correlations against NOAA raster data.}
    \label{fig:raleigh_compare_num_sensors}
\end{figure}

 \begin{figure}[H]
    \centering
    \includegraphics[width=0.7\textwidth]{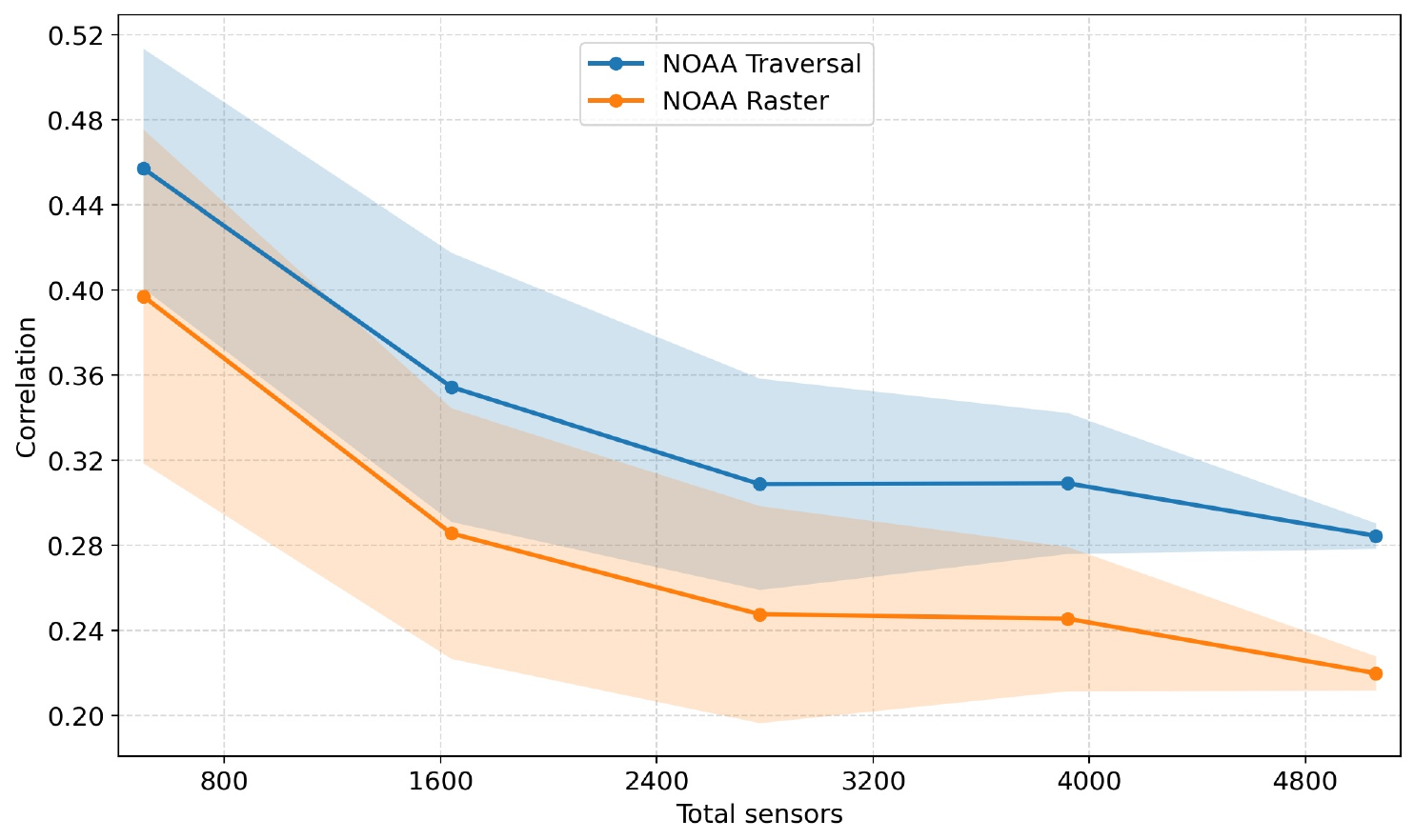}
    \caption{Predictive correlation with NOAA Urban Heat Island (UHI) campaign data as a function of total sensor count for Asheville on August 24, 2023 (6–7 A.M.). Correlations are computed between model-predicted block-group temperatures (using randomly selected subsets of sensors from that day) and (i) NOAA mobile traversal measurements (blue) and (ii) NOAA raster products (orange), both aggregated to block groups. Shaded bands indicate 95\% bootstrap confidence intervals. In contrast to all other cities, predictive agreement decreases with increasing sensor density.}
    \label{fig:asheville_compare_num_sensors}
\end{figure}

\section{Preferential sampling adjustment caveats}
\label{sec:appendix_ps_caveats}

We observed that the effects of preferential sampling correction were strongest for the afternoon data and weaker in the morning and evening, reflecting differences in both UHI intensity and day-to-day meteorological variability. In the morning, temperature variability is limited compared to later in the day; differences in heat are largely explained by covariates like LST. Under these conditions, the relationship between sampling density and the latent temperature field is weak, and preferential sampling adjustment provides little additional benefit. In the afternoon and evening, however, the UHI effect is more complex. Afternoon temperatures exhibit stronger spatial gradients but are also more sensitive to synoptic conditions such as wind, resulting in increased day-to-day variability. Evening temperatures show the strongest UHI effect due to differential heat retention, yet are more difficult to predict due to delayed thermal responses and advective effects. These factors reduce model stability and can obscure the gains from preferential sampling adjustment. Incorporating wind speed and direction into the modeling framework in the future may help address these limitations, particularly under atypical wind conditions that disrupt expected spatial temperature structure.

\section{The issue with NOAA validation}
\label{sec:appendix_noaa_validation}

NOAA raster products were generated by applying machine learning models to traversal point data using informative covariates (building volume, canopy cover, etc.) to produce spatially continuous estimates \parencite{Voelkel}. While these raster products are convenient for comparison, they are validated using random cross-validation, which fails to account for spatial dependence in the traversal data. When validation locations are geographically close to training points, this can result in overly optimistic estimates of raster accuracy. Thus, our models' correlations with NOAA raster products should be interpreted as informative but not definitive measures of predictive accuracy.

NOAA traversal data comprise dense, sequential temperature measurements collected along predefined routes by drivers or bicyclists \parencite{NOAA}. Although these data provide high spatial resolution along road networks, they are strongly autocorrelated in space and time and therefore do not represent independent samples. Aggregating traversal measurements to areal units implicitly assumes independence when the mean is calculated, leading to overestimation of effective sample size and underestimation of uncertainty. This issue is particularly pronounced when traversal routes cover only a small portion of the areal unit, making it difficult to infer temperatures at locations far from explored paths. Thus, correlations between model predictions and traversal-aggregated temperatures should also be interpreted cautiously. We additionally tested block kriging of traversal data to areal units as an alternative to mean-aggregation; however, correlations did not change much. We therefore report NOAA raster correlations in the main analysis.

\section{Commentary on county-specific income effects}
\label{sec:appendix_income_commentary}

We may also propose some hypotheses for why we observe certain trends in how the preferential sampling effect depends on income. Notably, in the map shown in Figure 3 of the manuscript, the strongest positive income effects appear in counties such as Halifax, Transylvania, and Rutherford, possibly reflecting affluent enclaves (e.g. lakefront or retirement communities) or rapidly developing suburban areas where stations are concentrated in higher-income tracts. In contrast, mountain counties such as Ashe and Yancey exhibit weak or negative slopes, where station placement in lower-income areas may be driven more by agricultural needs than by household socioeconomic status.

\section*{Acknowledgments}
We thank NOAA and the CAPA Heat Watch program for providing urban heat island campaign data used in validation.

\section*{Funding Statement}
This work was supported by Underwriters Laboratories, Chemical Insights Research Institute, and the National Science Foundation (Grant No.\ 2022040) through Duke University's aiM program. The funders had no role in study design, data collection and analysis, decision to publish, or preparation of the manuscript.

\section*{Competing Interests}
The authors declare no competing interests.

\section*{Data Availability Statement}
Data from Weather Underground are available at \url{https://wunderground.com}. Elevation data are available from the USGS National Elevation Dataset (\url{https://www.usgs.gov/the-national-map-data-delivery}). MODIS LST products can be accessed through Google Earth Engine (\url{https://earthengine.google.com}). Demographic data were obtained from the U.S.\ Census Bureau's American Community Survey (\url{https://www.census.gov/programs-surveys/acs}). NOAA urban heat island campaign data are available at \url{https://www.heat.gov}. All code used for analysis is available at \url{https://github.com/elliekkim/nc_temp_ps}.

\section*{Author Contributions}
Conceptualization: E.K.K., D.E.C. Methodology: E.K.K., Z.D.C., D.E.C. Software: E.K.K. Data curation: E.K.K., Z.D.C. Formal analysis: E.K.K. Visualization: E.K.K. Writing -- original draft: E.K.K. Writing -- review \& editing: E.K.K., Z.D.C., M.H.B., D.E.C. Supervision: M.H.B., D.E.C. Funding acquisition: M.H.B., D.E.C. All authors approved the final submitted draft.

\printbibliography

@article{GEE,
  title   = {Google {Earth Engine}: Planetary-scale geospatial analysis for everyone},
  author  = {Gorelick, N. and Hancher, M. and Dixon, M. and Ilyushchenko, S. and Thau, D. and Moore, R.},
  journal = {Remote Sensing of Environment},
  volume  = {202},
  pages   = {18--27},
  year    = {2017},
  doi     = {10.1016/j.rse.2017.06.031}
}

@book{Oke,
  title     = {Urban Climates},
  author    = {Oke, T. R. and Mills, G. and Christen, A. and Voogt, J. A.},
  publisher = {Cambridge University Press},
  year      = {2017}
}

@article{Khatana,
  title   = {Association of extreme heat with all-cause mortality in the contiguous {US}, 2008--2017},
  author  = {Khatana, Sameed A. M. and Werner, Rachel M. and Groeneveld, Peter W.},
  journal = {JAMA Network Open},
  volume  = {5},
  number  = {6},
  pages   = {e2212957},
  year    = {2022},
  doi     = {10.1001/jamanetworkopen.2022.12957}
}

@article{Minor,
  title   = {Rising temperatures erode human sleep globally},
  author  = {Minor, K. and Bjerre-Nielsen, A. and Jonasdottir, S. S. and Lehmann, S. and Obradovich, N.},
  journal = {One Earth},
  volume  = {5},
  number  = {5},
  pages   = {534--549},
  year    = {2022},
  doi     = {10.1016/j.oneear.2022.04.008}
}

@inproceedings{gardner2018gpytorch,
  title     = {{GPyTorch}: Blackbox matrix-matrix {Gaussian} process inference with {GPU} acceleration},
  author    = {Gardner, Jacob R. and Pleiss, Geoff and Bindel, David and Weinberger, Kilian Q. and Wilson, Andrew Gordon},
  booktitle = {Advances in Neural Information Processing Systems},
  volume    = {31},
  year      = {2018}
}

@inproceedings{paszke2019pytorch,
  title     = {{PyTorch}: An imperative style, high-performance deep learning library},
  author    = {Paszke, Adam and Gross, Sam and Massa, Francisco and Lerer, Adam and Bradbury, James and Chanan, Gregory and Killeen, Trevor and Lin, Zeming and Gimelshein, Natalia and Antiga, Luca and Desmaison, Alban and Kopf, Andreas and Yang, Edward and DeVito, Zachary and Raison, Martin and Tejani, Alykhan and Chilamkurthy, Sasank and Steiner, Benoit and Fang, Lu and Bai, Junjie and Chintala, Soumith},
  booktitle = {Advances in Neural Information Processing Systems},
  volume    = {32},
  year      = {2019}
}

@misc{ACS,
  author       = {{U.S. Census Bureau}},
  title        = {American {Community} {Survey} ({ACS})},
  year         = {2023},
  howpublished = {\url{https://www.census.gov/programs-surveys/acs}},
  note         = {Accessed: 2026-02-18}
}

@misc{NOAA,
  title        = {Urban {Heat} {Island} Mapping Campaigns},
  author       = {{National Integrated Heat Health Information System (NIHHIS)}},
  howpublished = {\url{https://www.heat.gov/pages/nihhis-urban-heat-island-mapping-campaign-cities}},
  note         = {Accessed: 2025-08-23}
}

@article{Diggle,
  title   = {Geostatistical inference under preferential sampling},
  author  = {Diggle, P. J. and Menezes, R. and Su, T.},
  journal = {Journal of the Royal Statistical Society: Series C (Applied Statistics)},
  volume  = {59},
  number  = {2},
  pages   = {191--232},
  year    = {2010},
  doi     = {10.1111/j.1467-9876.2009.00701.x}
}

@article{Voelkel,
  title   = {Towards systematic prediction of urban heat islands: Grounding measurements, assessing modeling techniques},
  author  = {Voelkel, J. and Shandas, V.},
  journal = {Climate},
  volume  = {5},
  number  = {2},
  pages   = {41},
  year    = {2017},
  doi     = {10.3390/cli5020041}
}

@article{calhoun2024refining,
  title   = {Refining citizen climate science: Addressing preferential sampling for improved estimates of urban heat},
  author  = {Calhoun, Zachary D. and Black, Marilyn and Bergin, Mike and Carlson, David},
  journal = {Environmental Science \& Technology Letters},
  volume  = {11},
  number  = {8},
  pages   = {845--850},
  year    = {2024}
}

@article{essd-13-4349-2021,
  title   = {{ERA5-Land}: A state-of-the-art global reanalysis dataset for land applications},
  author  = {Mu{\~n}oz-Sabater, J. and Dutra, E. and Agust{\'i}-Panareda, A. and Albergel, C. and Arduini, G. and Balsamo, G. and Boussetta, S. and Choulga, M. and Harrigan, S. and Hersbach, H. and Martens, B. and Miralles, D. G. and Piles, M. and Rodr{\'i}guez-Fern{\'a}ndez, N. J. and Zsoter, E. and Buontempo, C. and Th{\'e}paut, J.-N.},
  journal = {Earth System Science Data},
  volume  = {13},
  number  = {9},
  pages   = {4349--4383},
  year    = {2021},
  doi     = {10.5194/essd-13-4349-2021}
}

@inproceedings{titsias_2009,
  title     = {Variational learning of inducing variables in sparse {Gaussian} processes},
  author    = {Titsias, Michalis},
  booktitle = {Proceedings of the Twelfth International Conference on Artificial Intelligence and Statistics},
  pages     = {567--574},
  year      = {2009},
  publisher = {PMLR}
}

@article{KassRaftery1995,
  author  = {Kass, Robert E. and Raftery, Adrian E.},
  title   = {Bayes factors},
  journal = {Journal of the American Statistical Association},
  volume  = {90},
  number  = {430},
  pages   = {773--795},
  year    = {1995},
  doi     = {10.1080/01621459.1995.10476572}
}

@unpublished{calhoun_2025_spatiotemporal,
  title  = {Scalable and robust {Gaussian} processes for reanalysis of urban air temperature with crowdsourced meteorological data},
  author = {Calhoun, Zachary D. and Bergin, Mike and Carlson, David},
  note   = {Manuscript in preparation},
  year   = {2025}
}

@article{hsiao2025inverse,
  title   = {Inverse sampling intensity weighting for preferential sampling adjustment},
  author  = {Hsiao, Thomas W. and Waller, Lance A.},
  journal = {arXiv preprint arXiv:2503.05067},
  year    = {2025}
}

@article{damianou2013deepgaussianprocesses,
  title   = {Deep {Gaussian} processes},
  author  = {Damianou, Andreas C. and Lawrence, Neil D.},
  journal = {arXiv preprint arXiv:1211.0358},
  year    = {2013}
}

@misc{ECONet,
  title        = {{ECONet}: The {Environment} and {Climate} {Observing} {Network}},
  author       = {{North Carolina State Climate Office (NCSCO)}},
  howpublished = {\url{https://climate.ncsu.edu/econet/}},
  note         = {Raleigh, NC. Accessed: 2026-02-16}
}

@inproceedings{seabold2010statsmodels,
  title     = {Statsmodels: Econometric and statistical modeling with {Python}},
  author    = {Seabold, Skipper and Perktold, Josef},
  booktitle = {Proceedings of the 9th Python in Science Conference},
  pages     = {92--96},
  year      = {2010},
  editor    = {van der Walt, St{\'e}fan and Millman, Jarrod}
}

@article{berman1992approximating,
  author  = {Berman, Mark and Turner, T. Rolf},
  title   = {Approximating point process likelihoods with {GLIM}},
  journal = {Journal of the Royal Statistical Society: Series C (Applied Statistics)},
  volume  = {41},
  number  = {1},
  pages   = {31--38},
  year    = {1992},
  doi     = {10.2307/2347614}
}

@book{kingman1993poisson,
  author    = {Kingman, J. F. C.},
  title     = {Poisson Processes},
  publisher = {Oxford University Press},
  year      = {1993}
}

@incollection{reich_2012_AccountingDesignAnalysis,
  title     = {Accounting for design in the analysis of spatial data},
  author    = {Reich, Brian J. and Fuentes, Montserrat},
  booktitle = {Spatio-Temporal Design},
  editor    = {Mateu, Jorge and M{\"u}ller, Werner G.},
  edition   = {1},
  pages     = {131--141},
  publisher = {Wiley},
  year      = {2012},
  doi       = {10.1002/9781118441862.ch6}
}

@article{pati2011bayesian,
  author  = {Pati, Debdeep and Reich, Brian J. and Dunson, David B.},
  title   = {Bayesian geostatistical modelling with informative sampling locations},
  journal = {Biometrika},
  year    = {2011},
  volume  = {98},
  number  = {1},
  pages   = {35--48},
  doi     = {10.1093/biomet/asq067}
}

@article{guan_2008_ConsistentNonparametricIntensity,
  title   = {On consistent nonparametric intensity estimation for inhomogeneous spatial point processes},
  author  = {Guan, Yongtao},
  journal = {Journal of the American Statistical Association},
  volume  = {103},
  number  = {483},
  pages   = {1238--1247},
  year    = {2008},
  doi     = {10.1198/016214508000000526}
}

@article{bingham2019pyro,
  author  = {Bingham, Eli and Chen, Jonathan P. and Jankowiak, Martin and Obermeyer, Fritz and Pradhan, Neeraj and Karaletsos, Theofanis and Singh, Rohit and Szerlip, Paul A. and Horsfall, Paul and Goodman, Noah D.},
  title   = {Pyro: Deep universal probabilistic programming},
  journal = {Journal of Machine Learning Research},
  volume  = {20},
  pages   = {28:1--28:6},
  year    = {2019}
}

@article{santamouris_2020_HeatMitigationTechnologies,
  title   = {Heat mitigation technologies can improve sustainability in cities. {An} holistic experimental and numerical impact assessment of urban overheating and related heat mitigation strategies on energy consumption, indoor comfort, vulnerability and heat-related mortality and morbidity in cities},
  author  = {Santamouris, M. and Paolini, R. and Haddad, S. and Synnefa, A. and Garshasbi, S. and Hatvani-Kovacs, G. and Gobakis, K. and Yenneti, K. and Vasilakopoulou, K. and Feng, J. and Gao, K. and Papangelis, G. and Dandou, A. and Methymaki, G. and Portalakis, P. and Tombrou, M.},
  journal = {Energy and Buildings},
  volume  = {217},
  pages   = {110002},
  year    = {2020},
  doi     = {10.1016/j.enbuild.2020.110002}
}

\end{document}